\documentclass[10pt]{article}

\usepackage{amsthm,amsfonts,amsmath,amssymb}
\usepackage{pxfonts}
\usepackage{euscript}
\usepackage{hyperref}

\usepackage{xypic}
\usepackage{graphicx}
\usepackage{epstopdf} 

\newcommand\void[1]       {}

\theoremstyle{definition}

\newtheorem{thm}{Theorem}[section]

\newtheorem{conj}[thm]{Conjecture}
\theoremstyle{definition}
\newtheorem{defn}[thm]{Definition}
\newtheorem{exam}[thm]{Example}
\newtheorem{rem}[thm]{Remark}

\numberwithin{equation}{section}
\numberwithin{thm}{section}

\newcommand\nn             {\nonumber \\}
\newcommand\be            {\begin{equation}}
\newcommand\ee            {\end{equation}}
\newcommand\bea           {\begin{eqnarray}}
\newcommand\eea         {\end{eqnarray}}
\newcommand\bnu          {\begin{enumerate}}
\newcommand\enu          {\end{enumerate}}

\allowdisplaybreaks[1]

\newlength{\fighskip} \fighskip=2pt
\newlength{\figvskip} \figvskip=3pt

\newcommand{\pf}{\begin{proof}}
\newcommand{\epf}{\end{proof}}

\newcommand\Cb            {\mathbb{C}}

\newcommand\Rb            {\mathbb{R}}
\newcommand\Zb            {\mathbb{Z}}

\newcommand\CA           {\EuScript{A}}
\newcommand\CB           {\EuScript{B}}
\newcommand\CC           {\EuScript{C}}
\newcommand\CD           {\EuScript{D}}
\newcommand\CE          {\EuScript{E}}

\newcommand\CM          {\EuScript{M}}
\newcommand\CN         {\EuScript{N}}

\newcommand\CX         {\EuScript{X}}

\newcommand\CAs{{\EuScript{A}^\sharp}}
\newcommand\CBs{{\EuScript{B}^\sharp}}
\newcommand\CCs{{\EuScript{C}^\sharp}}
\newcommand\CDs{{\EuScript{D}^\sharp}}

\newcommand\CMs{{\EuScript{M}^\sharp}}

\newcommand\CXs{{\EuScript{X}^\sharp}}

 \DeclareMathOperator{\Hom}{Hom}

 \DeclareMathOperator{\Id}{Id}
 \DeclareMathOperator{\id}{id}

 \DeclareMathOperator{\ev}{ev}

 \DeclareMathOperator{\Mod}{Mod}

\newcommand{\rev}{\mathrm{rev}}

\newcommand{\one}{\mathbf1}
\newcommand\bk{\mathbf{k}}
\newcommand\bh{\mathbf{H}}

\newcommand\cl{\mathrm{bulk}}
\newcommand{\op}{\mathrm{bdy}}

\begin{document}

\begin{center} \LARGE
Gapless edges of 2d topological orders and enriched monoidal categories
\end{center}

\vskip 0.8em
\begin{center}
{\large
Liang Kong$^{a,b}$,\,
Hao Zheng$^{c}$\,
~\footnote{Emails:
{\tt  lkong@math.tsinghua.edu.cn, hzheng@math.pku.edu.cn}}}
\\[0.5em]
$^a$ Yau Mathematical Science Center\\
Tsinghua University, Beijing, 100084, China
\\[0.3em]
$^b$ Department of Mathematics \& Statistics\\
University of New Hampshire, Durham, 03824, USA
\\[0.3em]
$^c$ Department of Mathematics\\
Peking University, Beijing 100871, China
\end{center}

\vskip 1.7em

\begin{abstract}
In this work, we give a precise mathematical description of a chiral gapless edge of a 2d topological order (without symmetry). We show that the observables on the 1+1D world sheet of such an edge consist of a family of topological edge excitations, boundary CFT's and walls between boundary CFT's. These observables can be described by a chiral algebra and an enriched monoidal category. This mathematical description automatically includes that of gapped edges as special cases. Therefore, it gives a unified framework to study both gapped and gapless edges. Moreover, the boundary-bulk duality also holds for gapless edges. More precisely, the unitary modular tensor category that describes the 2d bulk phase is exactly the Drinfeld center of the enriched monoidal category that describes the gapless/gapped edge. We propose a classification of all gapped and chiral gapless edges of a given bulk phase. In the end, we explain how modular-invariant bulk rational conformal field theories naturally emerge on certain gapless walls between two trivial phases. 
\end{abstract}

\vspace{0.5cm} 
\tableofcontents
\vspace{0.5cm}


\section{Introduction}

It is well-known that the fusing-braiding properties of topological excitations in a 2d topological order (without symmetry) can be described by a unitary modular tensor category (UMTC) (see 
a review \cite[Appendix\,E]{kitaev}). A 2d topological order is uniquely determined by a pair $(\CC,c)$, where $\CC$ is the UMTC of topological excitations and $c$ is the chiral central charge. If the topological order $(\CC,0)$ has a gapped edge, the edge can be mathematically described by a unitary fusion category (UFC) $\CM$ such that the Drinfeld center $Z(\CM)$ of $\CM$ coincides with $\CC$ \cite{kk,fsv,anyon}. 
The fact that the bulk phase is uniquely determined by the gapped edge as its Drinfeld center is also called {\it boundary-bulk duality}. 

\smallskip
In general, 2d topological orders (without symmetry), such as quantum Hall systems, 
have topologically protected gapless edges \cite{halperin,wen1,moore-read,wen2} (see reviews \cite{wen2,wen3,wen4,nssfs} and references therein). A gapless edge is significantly richer than a gapped edge because gapless edge modes are described by 1+1D rational conformal field theories (RCFT's) \cite{bpz,moore-seiberg}, the mathematical structures of which are much richer than a UMTC \cite{moore-seiberg,turaev,klm,huang-mtc}. As far as we know, the precise mathematical description of a gapless edges is still not known. 

\smallskip
In the last twenty years, the mathematical theory of boundary-bulk (or open-closed) RCFT's has been successfully developed from at least three different perspectives (see the conformal-net approach  in \cite{longo-rehren,rehren1,rehren2,klm}, the 2+1D-TQFT approach in \cite{fffs,fs1,frs1,fjfrs} and the vertex-operator-algebra approach in \cite{huang-geo-voa,osvoa,ffa,kong-cardy}, and references therein). These mathematical developments have revealed a universal phenomenon: the mathematical structures of a boundary-bulk RCFT can be split into two parts. One part consists of a unitary rational chiral algebra $V$ (or a conformal net in the first approach), also called a unitary rational vertex operator algebra (VOA) (see for example \cite{ll,dong-lin}) in mathematics, such that the category $\Mod_V$ of $V$-modules is a UMTC \cite{huang-mtc}. The other part is a pure categorical structure containing certain algebras in $\Mod_V$ and its Drinfeld center $Z(\Mod_V)$. This suggests that it might be possible to describe a chiral gapless edge of a 2d topological order $(\CC,c)$ by a pair $(V,\CMs)$, where $V$ is a VOA and $\CMs$ is a purely categorical structure that can be constructed from $\Mod_V,\CC$ and perhaps additional categorical data. The main goal of this paper is to show that this is indeed possible. 

\medskip
In Section\,\ref{sec:voa}, we recall some basic facts of boundary-bulk CFT's. We explain that if a boundary-bulk CFT preserves a chiral symmetry given by a VOA $V$ on the boundary, then there is a very simple but equivalent categorical description of such a boundary-bulk RCFT as certain algebras in UMTC's. We summarize the results in a physics/mathematics dictionary at the end of this section. In Section\,\ref{sec:observables}, we explain that observables on a chiral gapless edge of a 2d topological order consist of a family of topological edge excitations, boundary CFT's and walls between these boundary CFT's. These boundary CFT's and walls are required to preserve a chiral symmetry given by a VOA $V$ such that $\Mod_V$ is a UMTC. This symmetry condition allows us to describe all boundary CFT's and walls equivalently by certain algebras and objects in $\CB:=\Mod_V$. As a consequence, all these observables organize themselves into a $\CB$-enriched monoidal category $\CXs$ (see Def.\,\ref{def:emc}). We denote such a gapless edge by the pair $(V,\CXs)$. In Section\,\ref{sec:can-edge}, we describe a canonical gapless edge $(V,\CBs)$ of a 2d bulk phase $(\CB,c)$, where the enriched monoidal category $\CBs$ is canonically obtained from $\CB$. The boundary-bulk duality holds in this case. Namely, the Drinfeld center $Z(\CBs)$ of $\CBs$ coincides with the UMTC $\CB$ \cite{kz2}. In Section\,\ref{sec:general-edges}, we explain that we can obtain a new gapless edge of a bulk phase $(\CC,c)$ by fusing a gapped domain wall $\CM$ between two bulk phases $(\CC,c)$ and $(\CB,c)$ with the canonical gapless edge $(V,\CBs)$ of $(\CB,c)$. This new edge can again be described by a pair $(V,\CMs)$, where $\CMs$ is a $\CB$-enriched monoidal category canonically constructed from the pair $(\CB,\CM)$. Moreover, the boundary-bulk duality holds, i.e. $Z(\CMs)=\CC$. This mathematical description of gapless edges (as pairs $(V,\CMs)$) automatically includes that of gapped edges as special cases. In this way, we have obtained a unified mathematical theory of gapless and gapped edges. This leads us to propose, in Section\,\ref{sec:classification}, a mathematical classification of all gapped and chiral gapless edges of a given bulk phase $(\CC,c)$. 
In Section\,\ref{sec:bcft}, we explain how a modular-invariant bulk CFT naturally emerges as a hom space in an enriched monoidal category describing certain gapless walls between two trivial 2d topological orders. We also discuss briefly 0d defects between edges. In Section\,\ref{sec:summary}, we give a summary and outlooks. In Appendix, we recall the mathematical definitions of various algebras in a UMTC and the definition of an enriched monoidal category.

\medskip
\noindent {\bf Acknowledgement}: We thank Maissam Barkeshli, Meng Cheng, Yi-Zhi Huang, Yuan-Ming Lu, Chetan Nayak, Xiao-Gang Wen and Yi-Zhuang You for very helpful discussions. LK is supported by the start fund from Tsinghua University. HZ is supported by NSFC under Grant No. 11131008.

\section{Basics of boundary-bulk CFT's} \label{sec:voa}

In this section, we briefly review the categorical description of various ingredients of boundary-bulk CFT's \cite{frs1,kong-cardy} (see a review  \cite{geometry} and references therein). The mathematically definitions of various algebras in UMTC's are given in Section\,\ref{sec:alg-umtc}. 

\medskip
The most important ingredient of a 2d CFT is a chiral algebra $V$ \cite{moore-seiberg}. It was defined rigorously in mathematics as a vertex operator algebra (VOA) (see for example \cite{ll}). It is a graded vector space $V=\oplus_{n=0}^\infty V_{(n)}$. The grading $n$ for $V_{(n)}$ is called the {\it conformal weight}, which is also an eigenvalue of the grading operator $L(0)$, i.e. $L(0) \cdot V_{(n)} = n \cdot V_{(n)}$. The {\it partition function} of $V$ is defined as follows   
$$
\chi_V(q) = \sum_{n=0}^\infty \dim V_{(n)} q^{n-\frac{c}{24}}, \quad\quad\quad \mbox{for $q\in \Cb^\times$}. 
$$ 
By the state-field correspondence, for each $\phi\in V$, there is a unique chiral field $\phi(z)$ associated to $\phi\in V$. The chiral field $\phi(z)$ depends on $z$ holomorphically. More precisely, $\phi(z)$ can be expanded as $\phi(z) = \sum_{n\in \Zb} \phi_n z^{-n-1}$, where $\phi_n$ is a linear operator that maps $V_{(m)}$ into $V_{(k-n-1+m)}, \forall m$ if $\phi\in V_{(k)}$. There is a distinguished weight 2 element $T\in V_{(2)}$. The chiral field $T(z)$ is the called the {\it energy-stress tensor} and can be expanded as follows:
$$
T(z) = \sum_{n\in \Zb} L(n) z^{-n-2}, 
$$ 
where $L(0)$ is the grading operator and $L(n), n\in \Zb$ generate a Lie algebra called {\it Virasoro algebra} defined by 
$$
[L(m), L(n)] = (m-n) L(m+n) + \delta_{m+n,0} \frac{m^3-m}{12} c \Id_V,  
$$
where $c$ is a complex number called the central charge of $V$. Chiral fields in $V$ have operator product expansions (OPE), i.e. 
$$
\phi(z_1) \psi(z_2) \sim \frac{ (\phi_k\psi)(z_2)}{(z_1-z_2)^{k+1}}  + \frac{ (\phi_{k-1}\psi)(z_2)}{(z_1-z_2)^{k}} + \cdots. 
$$
The OPE is commutative, i.e. $\phi(z_1) \psi(z_2) \sim \psi(z_2) \phi(z_1)$. In mathematics, this OPE and its properties were rigorously defined as the data and axioms of a VOA.

A simple $V$-module $W$ (or an irreducible representation of $V$) is again a graded vector space, i.e. $W=\oplus_{n=0}^\infty W_{(n)}$, where $L(0) \cdot W_{(n)} = (h_W + n) W_{(n)}$ for some $h_W \in \Cb$. For example, the Ising VOA $V$ has three simple $V$-modules, $\one, \psi, \sigma$, where $\one$ is just $V$ itself and $h_\one=0, h_\psi=\frac{1}{2}, h_\sigma=\frac{1}{16}$. The partition function of $W$ is defined by $\chi_W(q) = \sum_{n=0}^\infty \dim W_{(n)} q^{n+h_W-\frac{c}{24}}$. The set of all $V$-modules and $V$-module maps (linear maps that intertwine the $V$-actions) form the category of $V$-modules, denoted by $\Mod_V$.

The states (or chiral fields) in two different $V$-modules can be fused into the third $V$-module according to the so called {\it chiral vertex operators}\footnote{This notion was introduced by Moore and Seiberg \cite{moore-seiberg} for RCFT's, and was mathematically defined in \cite{fhl} by the name of {\it an intertwining operator}.}, which also have OPE. These OPE were conjectured in \cite{moore-seiberg} for RCFT's, and was mathematically proved in \cite{huang-JPAA,huang-ope-2}, and was shown to provide a monoidal structure on $\Mod_V$ \cite{hl1,hl2,hl3,hl4,huang-JPAA}. Namely, two $V$-modules $x$ and $y$ can be fused to give a new $V$-module $x\otimes y=y\otimes x$, which is a direct sum of simple $V$-modules. For example, for Ising VOA $V$, we have 
$$
\one \otimes \psi = \psi, \quad \one \otimes \sigma = \sigma, \quad \psi\otimes \sigma =\sigma, \quad \one = \one \otimes \one = \psi \otimes \psi, \quad  \sigma \otimes \sigma = 1\oplus \psi. 
$$ 
The fusion product $\otimes$ and the braidings $x\otimes y\xrightarrow{c_{x,y}} y\otimes x$ endow $\Mod_V$ with a braided monoidal structure. Moreover, when the VOA $V$ is unitary \cite{dong-lin} and rational \cite{huang-mtc}, the category $\Mod_V$ is a unitary modular tensor category (UMTC) \cite{huang-genus-1,huang-mtc,huang-mtc2}. The tensor unit $\one$ in $\Mod_V$ is nothing but $V$ itself. Each object $x$ in $\Mod_V$ has a dual object denoted by $x^\ast$ and the duality maps $u_x: \one \to x\otimes x^\ast$ and $v_x: x^\ast \otimes x \to \one$. For example, for Ising VOA $V$, all objects in $\Mod_V$ are self-dual, i.e. $\one=\one^\ast, \psi=\psi^\ast, \sigma=\sigma^\ast$. 

\medskip
In a boundary CFT \cite{cardy1,cardy2,cl}, the chiral fields on the 0+1D boundary (also called boundary fields) also have OPE \cite{cl,frs2}, which form a mathematical structure called {\it open-string vertex operator algebra} (OSVOA) \cite{osvoa}. A VOA is automatically an OSVOA. In general, an OSVOA $A$ is $\Rb$-graded $A=\oplus_{i\in \Rb} A_{(i)}$. An OSVOA $A$ always contains a VOA generated by the energy-stress tensor $T_A\in A_{(2)}$ and $T_A(r)=\sum_{n\in \Zb} L(n)r^{-n-2}$, where $r$ is the coordinate of the boundary and can be chosen to be  positive real numbers. In general, a boundary field $\phi(r)=\sum_{n\in \Rb} \phi_n r^{-n-1}$ can have non-integer powers of $r$, and the OPE of an OSVOA is not commutative. Therefore, an OSVOA can be viewed as a non-commutative generalization of a VOA. 

An OSVOA $A$ is called an OSVOA over $V$ if it is an extension of a VOA $V$ by $V$-modules and boundary fields in $A$ are all chiral vertex operators of $V$. This VOA $V$ should be viewed as the {\it chiral symmetry} of $A$. When the chiral symmetry $V$ is unitary and rational, $\Mod_V$ is not only a UMTC, but also a vertex tensor category \cite{hl1,huang-JPAA,huang-mtc2}, which allows us to reduce the complicated OPE structures in $A$ to a simple categorical structure: an algebra $A$ in the UMTC $\Mod_V$ \cite{osvoa} (conjectured in \cite{ko,fs1}). By definition, an algebra $A$ in $\Mod_V$ is just an object $A$ (i.e. a $V$-module), together with a multiplication morphism $m: A\otimes A \to A$ and a unit morphism $\iota: \one \to A$, i.e. a triple $(A,m,\iota)$, such that 
$$
m\circ (m\otimes \Id_A) = m\circ (\Id_A \otimes m), \quad\quad  m\circ (\iota \otimes \Id_A) = \Id_A = m \circ (\Id_A \otimes \iota). 
$$
The information of OPE is completely encoded in the morphism $m: A\otimes A \to A$; and that of the chiral symmetry is encoded in the morphism $\iota: \one\to A$ (i.e. an embedding $V\hookrightarrow A$) and the fact that both $A$ and $m$ are in $\Mod_V$. This miraculous simplification is the key to the success of the categorical classification of boundary-bulk RCFT's in \cite{frs1,fjfrs,kong-cardy,kr2}. Using this equivalence, it is very easy to construct OSVOA's over $V$. For example, let $x$ be a $V$-module, i.e. an object in $\Mod_V$. Then the internal hom $[x,x]=x\otimes x^\ast$, together with 
\be \label{eq:m-iota}
m: x\otimes x^\ast \otimes x\otimes x^\ast \xrightarrow{\Id_x \otimes v_x \otimes \Id_{x^\ast}} x\otimes x^\ast \quad\quad  and \quad\quad \iota: \one \xrightarrow{u_x} x\otimes x^\ast,
\ee
gives an algebra in $\Mod_V$, i.e. an OSVOA over $V$. 

\begin{rem}
If $A$ happens to be a VOA extension of $V$, then it is equivalent to a commutative algebra in $\Mod_V$ \cite{hkl} (conjectured in \cite{ko}), i.e. an algebra such that $m\circ c_{A,A} = m$. For example, $[\one,\one]=\one$ is the simplest commutative algebra in $\Mod_V$. 

\end{rem}

A boundary CFT $A_\op$ over $V$ is an OSVOA over $V$ equipped with a non-degenerate invariant bilinear form, which upgrade $A_\op$ to a symmetric Frobenius algebra in $\Mod_V$ \cite{kong-cardy} (see Def.\,\ref{def:ssfa}). 


\medskip
A bulk CFT contains bulk fields $\phi(z,\bar{z})$ that depend on both the holomorphic variable $z$ and the anti-holomorphic variable $\bar{z}$. Bulk fields also have OPE, which form a mathematical structure called a {\it full field algebra} \cite{ffa}. Let $U$ be a VOA, and let $\overline{V}$ be the same VOA as $V$ but consisting of only anti-chiral fields $\psi(\bar{z})$ for $\psi\in V$. The tensor product $U\otimes_\Cb \overline{V}$ with $\phi(z,\bar{z}) = u(z) \otimes_\Cb v(\bar{z})$ for $\phi=u\otimes_\Cb v\in U\otimes_\Cb \overline{V}$ gives an example of full field algebra. A full field algebra $A_\cl$ is called a full field algebra over $U\otimes_\Cb \overline{V}$ if it is an extension of $U\otimes_\Cb \overline{V}$ by objects in $\Mod_U\boxtimes \overline{\Mod_V}$ and all bulk fields are chiral vertex operators of the VOA $U\otimes_\Cb V$. $U$ is called the {\it chiral symmetry} of $A_\cl$ and $V$ is called the {\it anti-chiral symmetry} of $A_\cl$. 
A full field algebra over $U\otimes_\Cb \overline{V}$ is equivalent to a commutative algebra in $\Mod_U\boxtimes \overline{\Mod_V}$ \cite{kong-ffa}. A modular-invariant bulk CFT over $U\otimes_\Cb \overline{V}$ is a full field algebra over $U\otimes_\Cb \overline{V}$ equipped with a non-degenerate invariant bilinear form and a unique vacuum such that its genus-one correlation functions are all modular-invariant \cite{ffa-mod-inv}. It is equivalent to a Lagrangian algebra in $\Mod_U\boxtimes \overline{\Mod_V}$ \cite{kong-cardy,kr2} (see Def.\,\ref{def:lag-alg}). For example, the well-known charge-conjugate modular-invariant bulk CFT over $V\otimes_\Cb \overline{V}$ is given by the Lagrangian algebra $\oplus_i i\boxtimes i^\ast$ in $\Mod_V\boxtimes\overline{\Mod_V}$, where $i$ are simple objects in $\Mod_V$. Its partition function $\sum_i \chi_i(q) \chi_{i^\ast}(\bar{q})$ is modular invariant.

\medskip
A boundary-bulk CFT consists of a boundary CFT $A_\op$ and a modular-invariant bulk CFT $A_\cl$ satisfying some compatibility conditions \cite{kong-cardy,kr2} (see a different formulation in \cite{frs1,frs2,frs3}), one of which requires that the chiral symmetry of $A_\op$ coincides with both the chiral and anti-chiral symmetries of $A_\cl$ (see \cite[Def.\,1.25]{ocfa}). This condition implies that $A_\op$ is a boundary CFT over $V$ and $A_\cl$ is a bulk CFT over $V\otimes_\Cb \overline{V}$. If we include all compatibility conditions (such as the Cardy condition), then $A_\op$ must be a connected special symmetric Frobenius algebra (CSSFA) in $\Mod_V$ (see Def.\,\ref{def:ssfa} \& Remark\,\ref{rem:cssfa}), and $A_\cl$ must be the Lagrangian algebra (see Def.\,\ref{def:lag-alg}) given by the full center $Z(A_\op)$ of $A_\op$ in $\Mod_V\boxtimes\overline{\Mod_V}$ \cite{fjfrs,kr2}. For example, the internal hom algebra $[x,x]$ for $x\in \Mod_V$ defined in Eq.\,(\ref{eq:m-iota}) are CSSFA's in $\Mod_V$, and are boundary CFT's of the same charge-conjugate modular invariant bulk CFT $\oplus_i i\boxtimes i^\ast$, which is also the full center of $[x,x]$ and a Lagrangian algebra in $\Mod_V\boxtimes\overline{\Mod_V}$. It turns out that mapping CSSFA's in $\Mod_V$ to their full centers defines a one-to-one correspondence between the set of Morita classes of CSSFA's in $\Mod_V$ and that of Lagrangian algebras in $\Mod_V\boxtimes\overline{\Mod_V}$ \cite{kr1,kr2}. 

\medskip
We summarize results reviewed in this section in the following dictionary: 
\begin{center}
\begin{tabular}
[c]{|c|c|}\hline
Physical terminologies & Mathematical terminologies \\\hline\hline
a unitary rational chiral algebra & a unitary VOA $V$ s.t. $\Mod_V$ is a UMTC \\ \hline
chiral vertex operators & intertwining operators \\ \hline
boundary fields OPE & an open-string VOA (OSVOA) \\ \hline
boundary fields OPE & an OSVOA over $V$  \\
with the chiral symmetry given by $V$ &  = an algebra in $\Mod_V$ \\ \hline
a boundary CFT over $V$  &  a symmetric Frobenius algebra in $\Mod_V$  \\ \hline 
a modular-invariant bulk CFT &   a Lagrangian algebra $A_\cl$  \\
over $U\otimes_\Cb \overline{V}$ &  in $\Mod_U\boxtimes \overline{\Mod_V}$ \\  \hline \hline
boundary-bulk CFT over $V$ contains: &   \\ \hline
1. a boundary CFT over $V$  & a CSSFA $A_\op$ in $\CC:=\Mod_V$  \\ \hline
2. a modular invariant bulk CFT over $V\otimes_\Cb\overline{V}$ & a Lagrangian algebra $A_\cl$ in $\CC\boxtimes \overline{\CC}$ \\ \hline
3. boundary-bulk duality & $A_\cl=Z(A_\op)$, \\ 
 & where $Z(A_\op)$ is the full center of $A$ \\ \hline 
\end{tabular}
\end{center}

\section{Observables on a chiral gapless edge}  \label{sec:observables}
A 2d topological order is described by a UMTC $\CC$ and the chiral central charge $c$, i.e. a pair $(\CC,c)$. The chiral central charge is defined by the difference between the central charges of the right movers and the left movers, i.e. $c=c_R - c_L$. It modulo 8 equals to the topological central charge of $\CC$. In this work, we only study chiral gapless edges. Objects in $\CC$ are topological excitations. The tensor unit $\one_\CC$ of $\CC$ represents the trivial topological excitation. We denote the fusion product in $\CC$ by $\otimes$. The simplest UMTC is the category $\bh$ of finite dimensional Hilbert spaces. The pair $(\bh,0)$ describes the trivial 2d topological order.

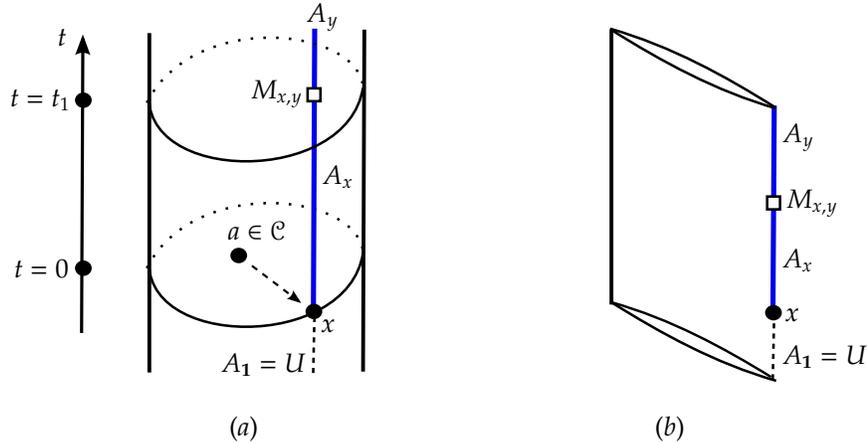
\begin{figure} 
$$
 \raisebox{-30pt}{
  \begin{picture}(130,130)
   \put(-20,8){\scalebox{0.6}{\includegraphics{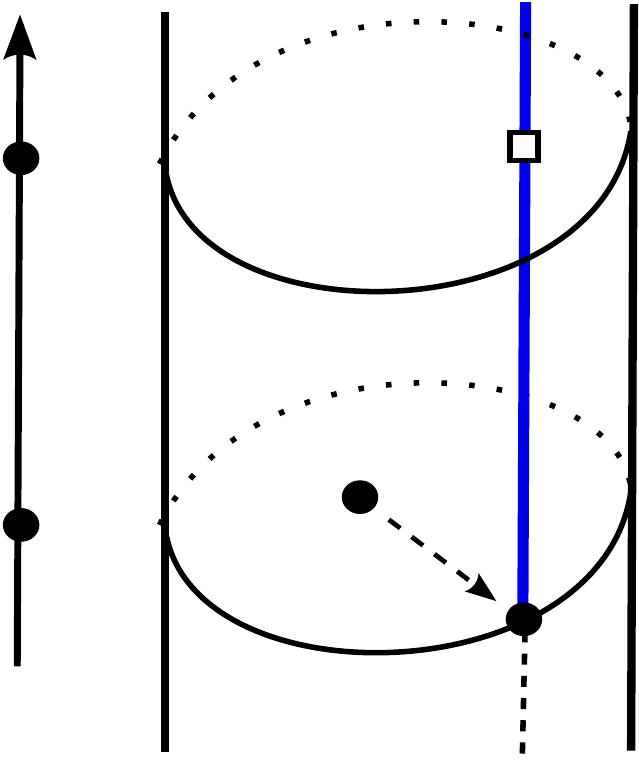}}}
   \put(-20,8){
     \setlength{\unitlength}{.75pt}\put(0,-83){
     \put(-30,133)  {$ t=0 $}
     \put(-32,219)  {$ t=t_1$}
     \put(-8, 250)  {$t$}
     \put(78,152)  {$ a \in \CC$}
     \put(126,180)  {$ A_x $}
     \put(118,262)  {$A_y$}
     \put(90,223)   {$M_{x,y}$}
     \put(75,85) {$A_{\one}=U$}
     \put(125, 105) {$x$}
     }\setlength{\unitlength}{1pt}}
  \end{picture}}
  \quad\quad\quad\quad\quad
 \raisebox{-30pt}{
  \begin{picture}(70,130)
   \put(0,5){\scalebox{0.6}{\includegraphics{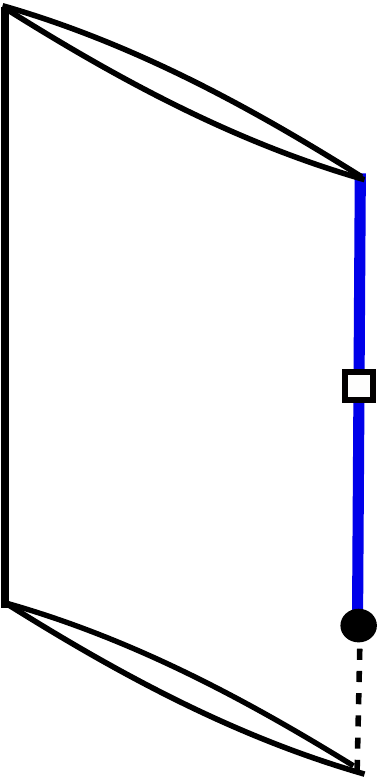}}}
   \put(0,5){
     \setlength{\unitlength}{.75pt}\put(0,-83){
     \put(88,142)  {$ A_x $}
     \put(88,205)  {$A_y$}
     \put(90,172)   {$M_{x,y}$}
     \put(88,93) {$A_{\one}=U$}
     \put(89, 115) {$x$}
     }\setlength{\unitlength}{1pt}}
  \end{picture}}
$$
$$
(a) \quad\quad\quad\quad\quad\quad\quad\quad\quad\quad\quad\quad\quad\quad\quad (b)
$$
\caption{The picture (a) depicts a 2d topological order $(\CC,c)$ on a 2-disk, together with a 1d gapless edge, propagating in time. When a topological bulk excitation $a\in \CC$ is moved to the edge at $t=0$, it creates a topological edge excitation $x$ or a boundary condition $M_x$ for the OSVOA $A_x$ living on the $t>0$ part of the world line. 
At $t=t_1>0$, the topological edge excitation $x$ is changed to another topological edge excitation $y$. This change creates a wall $M_{x,y}$ between $A_x$ and $A_y$. The picture (b) depicts the quasi-1+1D world sheet obtained by stretching the picture (a) along the dotted arrow from $a$ to $x$. 
}
\label{fig:cylinder}
\end{figure}

\medskip
Suppose that the 2d bulk phase $(\CC,c)$ is realized on an open 2-disk, and the edge of the 2-disk is gapless and chiral. We depict the 1+1D world sheet of the edge as a cylinder in Figure\,\ref{fig:cylinder} (a). The chiral fields associated to chiral edge modes live on the entire 1+1D world sheet. These chiral fields have OPE, which are described mathematically by a chiral algebra with the central charge $c$, or a VOA $U$ with the central charge $c$. In general, the category $\Mod_U$ of $U$-modules might not coincide with $\CC$ (see Section\,\ref{sec:general-edges}).

\medskip
This VOA $U$ is not the only observables on the gapless edge. In Figure\,\ref{fig:cylinder} (a), at $t=0$, a topological bulk excitation $a$ is moved to the edge, it creates a {\it topological edge excitation} (or a defect) labeled by $x$ on the edge at $t=0$. This excitation $x$ is different from those gapless excitations in the chiral edge modes, and should be viewed as certain super-selection sector, and is similar to a topological edge excitation on a gapped edge \cite{kk,icmp}). There are chiral fields (also called defect fields or chiral vertex operators) living on $x$ at $t=0$ \cite{wen1,wenwu1,wwh}. They form a vector space $M_x$. In general, non-trivial topological bulk excitations might condense on the edge, and there might be more topological edge excitations than those from the bulk.

Note that the topological edge excitation $x$ is also a wall between the $t>0$ part of the world line (the blue line in Figure\,\ref{fig:cylinder} (a)) and $t<0$ part. The chiral fields on the $t>0$ part of the world line supported on $x$ are potentially different from those in $U$. 
These chiral fields must also have OPE. Mathematically, this OPE structure of chiral fields in $A_x$ alone forms an OSVOA (see Section\,\ref{sec:voa}). We denote the trivial topological edge excitation by $\one$. When $x=\one$, it is clear that $A_\one=U$. We will argue that this OSVOA $A_x$ is one of the boundary CFT's of a modular invariant bulk CFT. Indeed, stretching the cylinder in Figure\,\ref{fig:cylinder} (a) along the dotted arrow from $a$ to $x$, we obtain the quasi-1+1D world sheet with two boundaries, one of which is the 0+1D world line supported on $x$, as depicted in Figure\,\ref{fig:cylinder} (b). Chiral fields on this boundary $A_x$ (and other data $A_y$ and $M_{x,y}$ which will be explained later) remain the same during this process. It is clear that both chiral and anti-chiral fields live on this quasi-1+1D world sheet. It was shown in \cite[Sec.\,2.1]{cz} that they form a modular invariant bulk CFT $A_{\mathrm{bulk}}$. This fact was also emphasized in \cite{rz,levin} as a consequence of the following no-go theorem: {\it any 1+1D conformal field theory realized by a 1d lattice Hamiltonian model is modular invariant}. Therefore, the OSVOA's $A_x$ is one of the boundary CFT's of the modular invariant bulk CFT $A_{\mathrm{bulk}}$. 

\begin{rem}
We have argued on the physical level of rigor that $A_x$ is a boundary CFT of a modular invariant bulk CFT. Mathematically, however, it is not a clear statement because there are different mathematical definitions of a boundary-bulk CFT. Here we assume that a boundary-bulk CFT satisfies Segal's type of axioms of an open-closed CFT of all genera \cite{huang-cft,kong-cardy}, including modular invariance condition, Cardy condition, etc. In other words, we propose the following conjecture (as a generalized ``no-go theorem''): {\it A 1+1D boundary-bulk conformal field theory realized by a 1d lattice Hamiltonian model with boundaries should satisfy the mathematical axioms of an open-closed CFT of all genera}. 
\end{rem}

\begin{rem}
Although using boundary CFT's to study 0d defects or impurities in other condensed matter systems has a long history \cite{al1,al2}, we were advised by referees that those condensed matter systems are quite different and can not be used to justify the appearance of boundary CFT's here. We are not aware of any earlier works that have mentioned or studied the boundary CFT's on the gapless edges of 2d topological orders. That chiral vertex operators live on the defect $x$ was known in 90's (see \cite{wenwu1,wwh}). But it is far from enough to imply that $A_x$ is a boundary CFT. See Remark\,\ref{rem:chiral-cft} for a further discussion. 
\end{rem}

\begin{rem}
We will come back in Section\,\ref{sec:bcft} to give a mathematically precise description of this stretching (or dimensional reduction) process and to identify precisely which modular invariant bulk CFT is obtained at the end of this process.
\end{rem}

\begin{figure} 
$$
 \raisebox{-50pt}{
  \begin{picture}(100,120)
   \put(-60,8){\scalebox{0.6}{\includegraphics{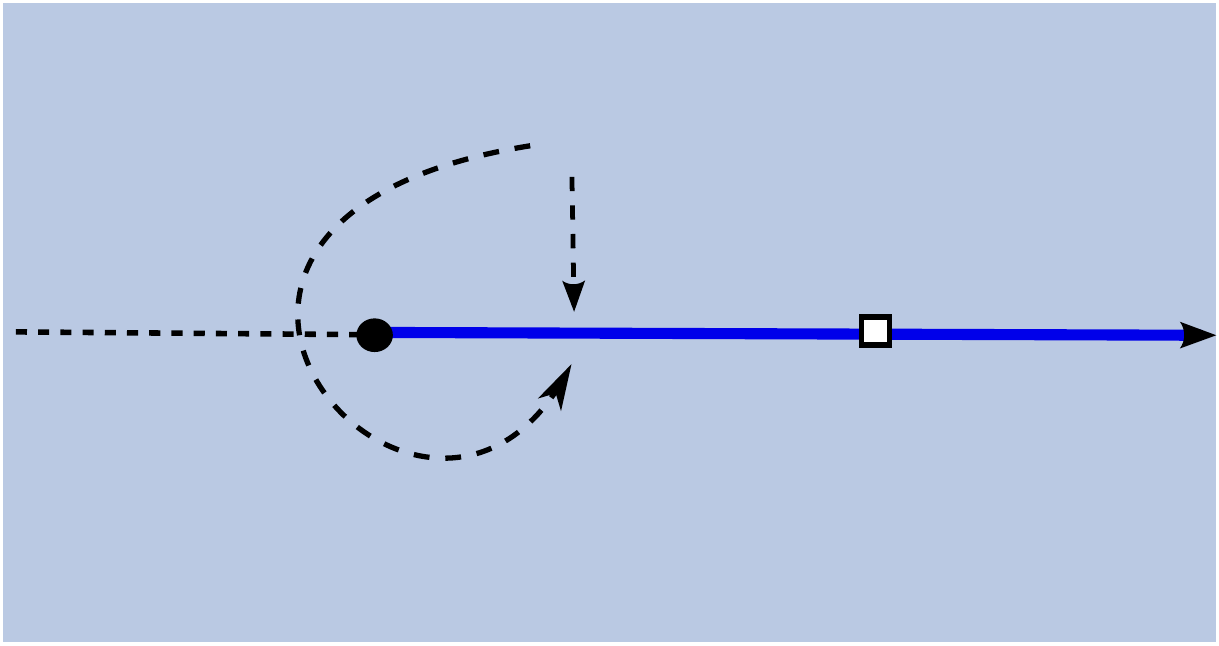}}}
   \put(-60,8){
     \setlength{\unitlength}{.75pt}\put(-23,0){
     \put(108,57)  {$ t=0 $}
     \put(295,57)  {$t$}
     \put(102,110)  {\scriptsize $\gamma_2$}
     \put(144,95)  {\scriptsize $\gamma_1$}
     \put(152,113) {\scriptsize $\times$}
     \put(150,125) {$\phi(z)\in V\subset U$}
     \put(175,58)   {\scriptsize $\Psi(t)\in A_x$}
     \put(175,70) {\scriptsize $\times$}
     \put(260,57)   {$A_y$}
     \put(220,82)   {\scriptsize $\Phi(t_1)\in M_{x,y}$}
     \put(223,58)   {$t_1$}
     \put(107, 80) {$x$}
     }\setlength{\unitlength}{1pt}}
  \end{picture}}
$$
\caption{This picture depicts the fusion of the chiral fields $\phi(z)$ in $U$ into those in $A_x$ along different paths. Chiral fields $\Psi(t)$ in the boundary CFT $A_x$ (or $A_y$) are restricted on the world line (the $t$-axis), which is also the blue line in Figure\,\ref{fig:cylinder}. Defect fields $\Phi(t_1)$ in $M_{x,y}$ are restricted on the point $t=t_1$. 
}
\label{fig:bcft}
\end{figure}
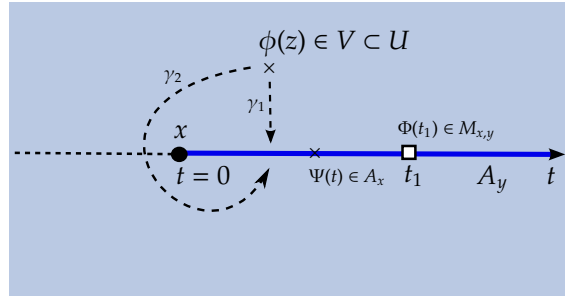

In order to be a well-defined boundary CFT, this OSVOA $A_x$ is required to satisfy certain compatibility conditions with $U$. More precisely, fusing chiral fields in $U$ into those in $A_x$ along a path $\gamma$ defines an OSVOA map $\iota_\gamma: U\to A_x$ (see Figure\,\ref{fig:bcft}). The minimal symmetry requirement for $A_x$ to be a consistent boundary CFT is that $\iota_\gamma$ should preserve the conformal symmetry \cite{cardy1}. More precisely, let $\langle T_U\rangle$ and $\langle T_{A_x}\rangle$ be the sub-VOA's in $U$ and $A_x$ generated by the energy-stress tensors $T_U\in U$ and $T_{A_x}\in A_x$, respectively. Preserving the conformal symmetry means that $\iota_\gamma|_{\langle T_U\rangle}: \langle T_U\rangle \to \langle T_{A_x}\rangle$ is a VOA isomorphism and independent of paths. More generally, one can require $\iota_\gamma$ to preserve a larger chiral symmetry given by a sub-VOA $V$ of $U$, i.e. $\iota_\gamma|_V: V \hookrightarrow A_x$ is an injective OSVOA homomorphism and independent of paths (see for example \cite{osvoa,ocfa}). This independence-of-path condition implies that $V$ is the chiral symmetry of $A_x$, or equivalently, $A_x$ must be a boundary CFT over $V$ (recall Section\,\ref{sec:voa}). At the same time, the bulk CFT $A_{\mathrm{bulk}}$ obtained in Figure\,\ref{fig:cylinder} (b) must be a bulk CFT over $V\otimes_\Cb \overline{V}$. This chiral symmetry $V$, or the $V$-symmetry, is an important data in describing the gapless edge. We assume that $V$ is unitary rational so that $\Mod_V$ is a UMTC. As we have discussed in Section\,\ref{sec:voa}, $A_x$ must be a connected special symmetric Frobenius algebra (CSSFA) in $\Mod_V$ (see Def.\,\ref{def:ssfa} \& Remark\,\ref{rem:cssfa}).

It is also possible to change a topological edge excitation $x$ to another $y$ on the same world line at $t=t_1>0$ as depicted in both Figure\,\ref{fig:cylinder} (a) and Figure\,\ref{fig:bcft}. For example, one can move a topological bulk excitation $b \in \CC$ to the world line at $t=t_1$ to give $y=b\otimes x$ in Figure\,\ref{fig:cylinder} (a). This process creates a wall (at $t=t_1$) between two boundary CFT's $A_x$ (on $\{0<t<t_1\}$) and $A_y$ (on $\{t>t_1\}$). From Figure\,\ref{fig:cylinder} (b), it is clear that the OSVOA's $A_x$ and $A_y$ are two boundary CFT's of the same bulk CFT $A_{\mathrm{bulk}}$, and are CSSFA's in $\Mod_V$. 
The defect fields on the wall are a special kind of chiral vertex operators called {\it boundary condition changing operators} \cite{cardy2}. They form a vector space $M_{x,y}$. It is clear that we should have $M_{\one,x}=M_x$ and $M_{x,x}=A_x$. Similar to $A_x$, the wall $M_{x,y}$ should also preserve the $V$-symmetry. This condition means that $M_{y,x}$ must be an object in $\Mod_V$. Moreover, defect fields in $M_{x,y}$ can be fused with those in $M_{y,z}$ to give defect fields in $M_{x,z}$.  This fusion should commute with the $V$-actions, i.e. an intertwining operator of $V$. Therefore, it can be described by a morphism $M_{y,z}\otimes M_{x,y} \to M_{x,z}$ in $\Mod_V$ \cite{hl1}. When $x=y=z$, this morphism is nothing but the multiplication morphism $A_x \otimes A_x \to A_x$ that defines the algebra structure on $A_x$. 

\begin{rem} \label{rem:U-non-local}
In general, $V\subsetneq U$ (see Remark\,\ref{rem:VM-V}). According to \cite{hkl}, the VOA $U$ (i.e. $A_\one$) is a commutative CSSFA in $\Mod_V$. Note that $U$ also acts on $M_{x,y}$. So $M_{x,y}$ is a left (or right) $U$-module. When $M_{x,y}=A_x$, $U$ acting on $A_x$ via two different paths $\gamma_1$ and $\gamma_2$ in Figure\,\ref{fig:bcft}. But $A_x$ is not necessarily a local $U$-module (\cite{bek,ko,frs1}) unless $V=U$ or $A_x=U$. According to \cite{hkl}, a VOA-module over $U$ is necessarily a local $U$-module in $\Mod_V$. Therefore, $M_{x,y}$ is not a VOA-module over $U$ in general. 
\end{rem}

In summary, all the observables on 1+1D world sheet of a gapless edge of a given 2d bulk phase $(\CC,c)$ can be described by a pair $(V,\CXs)$, where $\CXs$ is a categorical structure: 
\begin{itemize}
\item objects in $\CXs$ are topological edge excitations;  
\item for each pair $x,y$ objects in $\CXs$, there is a space $\hom_{\CXs}(x,y):=M_{x,y}$ which is an object in $\Mod_V$;  
\item an identity map $V = \one_{\Mod_V} \to M_{x,x}=A_x$ is a morphism in $\Mod_V$ defined by the canonical embedding $V\hookrightarrow A_x$;  
\item a composition map $M_{y,z}\otimes M_{x,y} \to M_{x,z}$ is a morphism in $\Mod_V$,
\end{itemize}
satisfying some natural physical properties, such as the unit property of the identity map and the associativity of the composition map. As a result, this categorical structure $\CXs$ is nothing but a category enriched in $\Mod_V$, or an $\Mod_V$-enriched category \cite{kelly} (see Def.\,\ref{def:en-cat}). 

\begin{figure} 
$$
 \raisebox{-70pt}{
  \begin{picture}(80,165)
   \put(-30,8){\scalebox{0.6}{\includegraphics{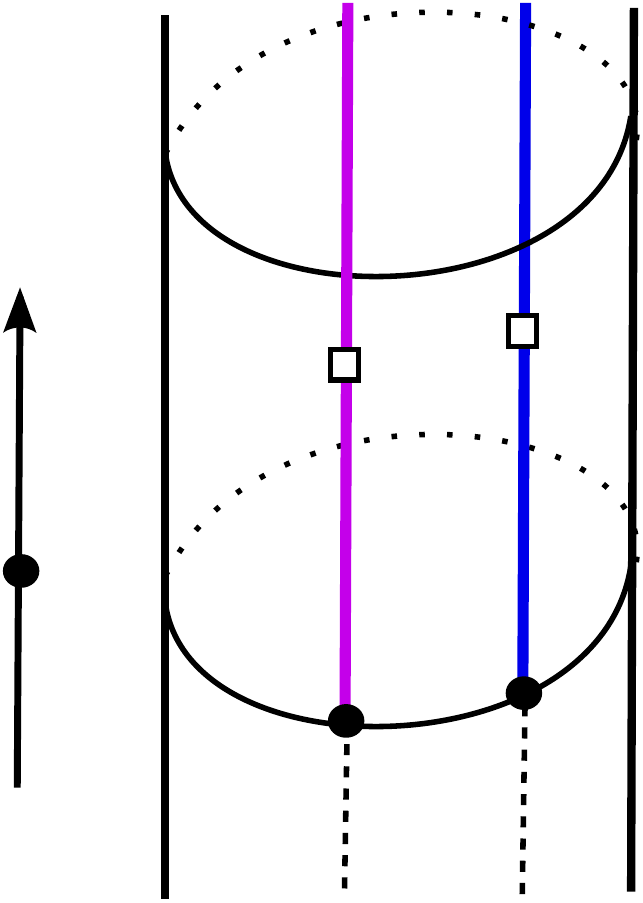}}}
   \put(-30,8){
     \setlength{\unitlength}{.75pt}\put(0,-70){
     \put(-30,140)  {$ t=0 $}
     \put(-10,180)  {$t$}
     \put(101,145)  {$ A_x $}
     \put(89,200)  {$ M_{x,y} $}
     \put(113,285)  {$A_y$}
     \put(43,192)  {$M_{x',y'}$}
     \put(59,139)  {$A_{x'}$}
     \put(73,285)  {$A_{y'}$}
     \put(65,75) {$U$}
     \put(125, 105) {$x$}
     \put(82,98)  {$x'$}
     }\setlength{\unitlength}{1pt}}
  \end{picture}}
\quad\quad\quad\quad \Rightarrow \quad\quad\quad 
 \raisebox{-70pt}{
  \begin{picture}(100,165)
   \put(10,8){\scalebox{0.6}{\includegraphics{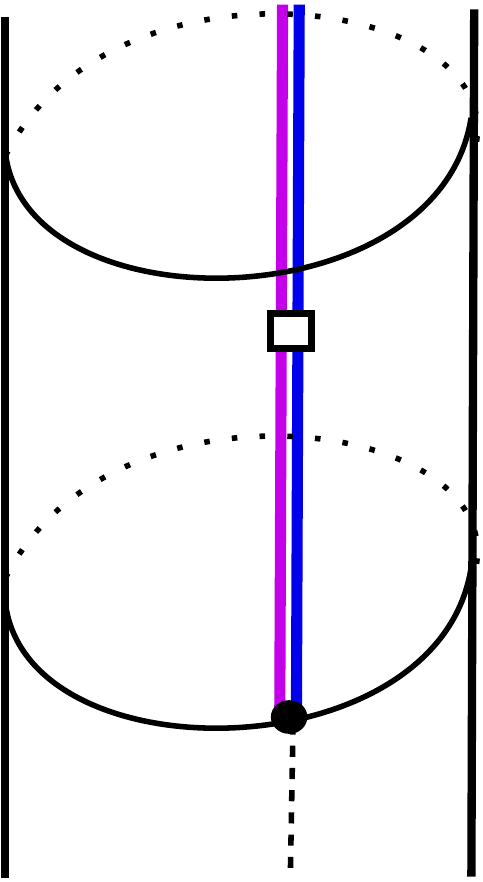}}}
   \put(10,8){
     \setlength{\unitlength}{.75pt}\put(-40,-75){
     \put(58,198)  {$M_{x'x,y'y}$}
     \put(76,145)  {$A_{x'x}$}
     \put(98,285)  {$A_{y'y}$}
     \put(93,80) {$U$}
     \put(112,102)  {$x'x$}
     }\setlength{\unitlength}{1pt}}
  \end{picture}}  
$$
$$
(a) \quad\quad\quad\quad\quad\quad\quad\quad\quad\quad\quad\quad\quad\quad\quad\quad
(b)
$$
\caption{This picture depicts how to fuse horizontally (on the same time slide) two topological edge excitations (or boundary conditions of boundary CFT's) $x$ and $y$, together with boundary CFT's $A_x$, $A_y$, $A_{x'}$, $A_{y'}$ and walls $M_{x,y}$, $M_{x',y'}$. For convenience, we abbreviate $x'\otimes x$ to $x'x$ in the picture. 
}
\label{fig:fusion}
\end{figure}
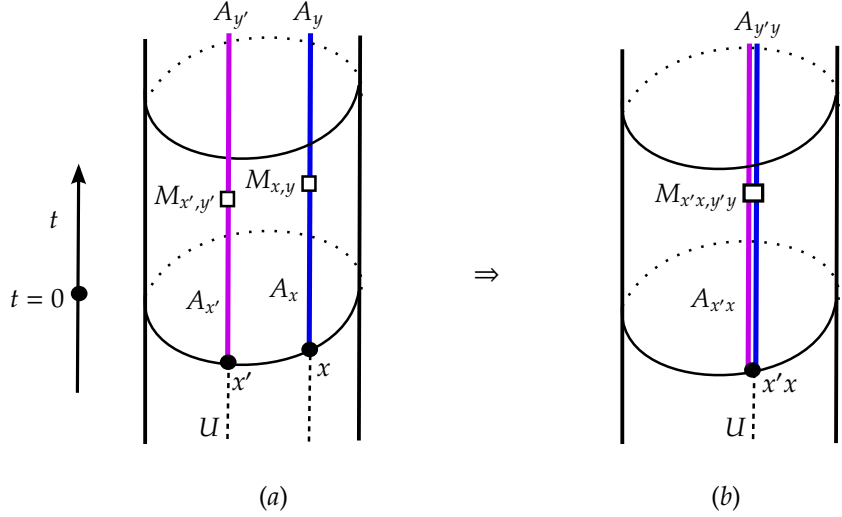

\smallskip
The last piece of structure is the fusion (on a time slide) of two topological edge excitations $x'$ and $x$, denoted by $x'\otimes x$ as depicted in Figure\,\ref{fig:fusion}. It automatically provides a fusion between observables on two parallel world lines. This fusion provides   
\begin{itemize}
\item a morphism $M_{x',y'}\otimes M_{x,y} \to M_{x'\otimes x, y'\otimes y}$ in $\Mod_V$ for objects $x,y,x',y'$ in $\CXs$, 
\end{itemize}
satisfying some natural properties. This fusion structure upgrades $\CXs$ to an $\Mod_V$-enriched monoidal category \cite{MP} (see Def.\,\ref{def:emc}).

\begin{rem}
Actually, the physics observables demand more structures than just an enriched monoidal category. For example, the topological edge excitations should be semisimple and have duals; the morphism $M_{x',y'}\otimes M_{x,y} \to M_{x'\otimes x, y'\otimes y}$ should factor through $M_{x',y'}\otimes_U M_{x,y} \simeq M_{x'\otimes x, y'\otimes y}$ (see Eq.\,(\ref{eq:tensor-product})), etc. These will lead us too far to the mathematical details. We will postpone the discussion of them to \cite{kz3}. 
\end{rem}

\begin{rem} \label{rem:bb-duality}
The boundary-bulk duality\footnote{It is related to but different from the bulk-edge correspondence (see \cite{moore-read,gww,sv,lwww} and references therein).} for topological orders in arbitrary dimensions was proved formally in \cite{kong-wen-zheng-2} under some natural assumptions. It says that the bulk topological order should be given by the center of the boundary phase regardless how we describe the bulk and the boundary phases mathematically. This formal proof also works for the cases in which the boundary phase is gapless \cite[Remark\,5.7]{kong-wen-zheng-2}. Therefore, if the enriched monoidal category $\CXs$ indeed gives a mathematical description of the gapless edge, then we expect that the Drinfeld center $Z(\CXs)$ of the enriched monoidal category $\CXs$, a notion which was recently introduced in \cite[Def.\,2.1]{kz2}, gives exactly the UMTC $\CC$, i.e. $Z(\CXs)=\CC$. 
\end{rem}

\section{A canonical chiral gapless edge} \label{sec:can-edge}

Let $\CB$ be a UMTC such that there exist a rational VOA $V$ of central charge $c$ such that $\Mod_V=\CB$. We denote its tensor unit by $\one_\CB$. The creation and annihilation of a particle-antiparticle pair are described by the duality morphisms 
\be \label{eq:duality-maps}
u_x: \one_\CB \to x\otimes x^*, \quad \quad \quad v_x: x^* \otimes x \to \one_\CB.
\ee 
We denote the braiding isomorphisms by $c_{x,y}: x\otimes y\to y\otimes x$ for $x,y\in \CB$.
We use $\overline{\CB}$ to denote the UMTC that is the same monoidal category as $\CB$ but equipped with the braidings given by the anti-braidings in $\CB$.  

\medskip
In this section, we focus on the simplest gapless edge of $(\CB,c)$, called the {\it canonical gapless edge}. We describe all the observables on this canonical gapless edge below. 
\begin{itemize}
\item Topological edge excitations are all obtained from moving topological bulk excitations to the edge. They are labeled by objects in $\CB$. 
\item $\CB=\Mod_V$ for a rational VOA $V$. 
\item $V=U$, namely, all the boundary CFT's $A_x$ and wall between them $M_{x,y}$ are required to preserve the largest chiral symmetry $V$. As a consequence,  $M_{x,y}$ are objects in $\CB$. 
\item $M_x=x$ as a $V$-module. For each $x\in \CB$, the boundary CFT $A_x$ is given by the internal hom $[x,x]:=x\otimes x^\ast$, which is a CSSFA in $\CB$ \cite{fffs,frs1,kong-cardy}. Its algebraic structures are defined in Eq.\,(\ref{eq:m-iota}). When $x=\one_\CB$, we have $A_{\one_\CB}=\one_\CB=V$. 

\item $M_{x,y}=[x,y]=y\otimes x^\ast$ \cite{ffrs}. In particular, $M_{\one_\CB,x}=M_x=[\one_\CB,x]=x$. 

\item Defect fields in $[x,y]$ can be fused with those in $[y,z]$ to give defect fields in $[x,z]$. This amounts to a morphism $[y,z] \otimes [x,y] \to [x,z]$ in $\CB$, which is defined as follows: 
\be \label{eq:composition}
[y,z] \otimes [x,y] = z\otimes y^\ast \otimes y \otimes x^\ast \xrightarrow{\Id_z \otimes v_y \otimes \Id_{x^\ast}} z\otimes x^\ast = [x,z]. 
\ee
It is clear that $[x,y]$ is automatically a $[y,y]$-$[x,x]$-bimodule.
\item The fusion of topological edge excitations (on the same time slide) depicted in Figure\,\ref{fig:fusion} demands a morphism $M_{x',y'}\otimes M_{x,y} \to M_{x'\otimes x, y'\otimes y}$ in $\CB$. It is defined as follows: 
\be \label{eq:tensor-product}
[x',y']\otimes [x,y]  = y' \otimes x'^* \otimes y \otimes x^* \xrightarrow{\Id_{y'} \otimes c_{x'^*, y\otimes x^*}} (y' \otimes y) \otimes (x'\otimes x)^* = [x'\otimes x, y'\otimes y].
\ee

\end{itemize}
This canonical edge of $(\CB,c)$ is the most studied edge in physics. But our description of the complete set of observables on this edge, especially the fusion of defects fields in (\ref{eq:composition}) and (\ref{eq:tensor-product}), in terms of internal homs is new. 

\begin{rem} \label{rem:xx-zz}
Recall that CSSFA's $[x,x]$ for $x\in \CB$ are all Morita equivalent to the tensor unit $\one_\CB$, and are boundary CFT's of the same charge-conjugate modular invariant bulk CFT given by the full center of $\one_\CB$, i.e. $Z(\one_\CB)=\oplus_i i\boxtimes i^\ast$, in $\CB\boxtimes\overline{\CB}$. In general, there are more CSSFA's in $\CB$ (not Morita equivalent to $[x,x]$) that can occur on a different gapless edge (see Remark\,\ref{rem:cl-cft}). 
\end{rem}

These observables on the canonical edge of $(\CB,c)$ can be summarized by a pair $(V,\CBs)$, where $\CBs$ is a categorical structure:
\begin{itemize}

\item An object in $\CBs$ is a topological edge excitation, i.e. an object $x$ in $\CB$; 

\item the hom space $\hom_\CBs(x,y)=[x,y]=y\otimes x^*$;  

\item a distinguished morphism $\id_x: \one_\CB \to [x,x]=x\otimes x^*$ defined by $u_x: \one_\CB \to x\otimes x^*$. 

\item a composition map $[y,z] \otimes [x,y] \to [x,z]$ defined by Eq.\,(\ref{eq:composition}). 

\item a morphism: $[x',y']  \otimes [x,y]  \to [x'\otimes x, y'\otimes y]$ defined by Eq.\,(\ref{eq:tensor-product}). 

\end{itemize}
It was proved in \cite{MP} that this categorical structure $\CBs$ is a $\CB$-enriched monoidal category. We denote this canonical edge by the pair $(V,\CBs)$. It is explained in Example\,\ref{rem:canonical-construction} that this enriched monoidal category $\CBs$ is exactly the one obtained from the pair $(\CB, \CB)$ via the canonical construction. Therefore, we can also denote $\CBs$ by the pair $(\CB,\CB)$, i.e. $\CBs=(\CB,\CB)$, where the second $\CB$ is viewed as a UFC equipped with the unitary braided monoidal functor $\phi_\CM: \overline{\CB} \to \overline{\CB}\boxtimes \CB=Z(\CB)$.

\begin{rem} \label{rem:en-in-H}
The notion of an enriched monoidal category is a generalization of the usual notion of a monoidal category. For example, an ordinary UFC $\CM$ can be viewed as the $\bh$-enriched monoidal category obtain from the pair $(\bh, \CM)$ via the same canonical construction given in Example\,\ref{rem:canonical-construction}, i.e. $\CM=(\bh,\CM)$. 
\end{rem}

Our categorical description of the canonical gapless edge need pass an important consistence check: boundary-bulk duality (recall Remark\,\ref{rem:bb-duality}). We expect the Drinfeld center $Z(\CBs)$ of $\CBs$ to coincide with $\CB$ as UMTC's. Indeed, we introduced the notion of Drinfeld center of an enriched monoidal category in \cite[Def.\,2.1]{kz2} and proved that $Z(\CBs)=\CB$ \cite[Cor.\,2.5]{kz2}. 

\begin{rem} \label{rem:chiral-cft}
It is generally accepted that a chiral gapless edge of a 2d topological order corresponds to the boundary of a 2+1D TQFT, and there are also convincing arguments that such a boundary supports a {\it chiral CFT}, which consists of a VOA $V$ and conformal blocks (i.e. hom spaces in $\Mod_V$). On the gapless edge, however, one only see chiral fields instead of the hom spaces in $\Mod_V$. As a consequence, the hom spaces are replaced by internal homs, and $\Mod_V$ is replaced by enriched monoidal category $\CBs$. This change is not contradicting to the statement ``such a boundary supports a {\it chiral CFT}'' because $\CBs$ is canonically obtained from $\CB$, and $\CB$ can be recovered from $\CBs$ as its underlying category (see Definition\,\ref{def:underlying}). But this change is not only physically natural but also crucial for the boundary-bulk duality to hold because $\CB$ describes a gapped edge of a double-layered bulk, i.e. $Z(\CB)=\CB\boxtimes \overline{\CB}$, while $\CBs$ describes a gapless edge of $(\CB,c)$, i.e. $Z(\CBs)=\CB$. 
\end{rem}

The conclusion of this section is that the complete mathematical description of the canonical gapless edge of $(\CB,c)$ is given by a pair $(V, \CBs)$, where $V$ is a VOA such that $\Mod_V=\CB$, and $\CBs$ is the $\CB$-enriched monoidal category obtained from the pair $(\CB,\CB)$ via the canonical construction given in Remark\,\ref{rem:canonical-construction}.

\section{General chiral gapless edges} \label{sec:general-edges}
Similar to gapped edges, in general, there are more than one chiral gapless edges  for a given bulk phase $(\CC,c)$ (see for example \cite{pmn,ccbcn,ccmnpy,bn}). 

\medskip
Similar to the fusion of gapped domain walls \cite{kk,fs2,lww,hw,ai,kawahigashi}, 
we can obtain a new gapless edge of the 2d bulk phase $(\CC,c)$ by fusing a canonical edge $(V,\CBs)$ of $(\CB,c)$ with
a gapped wall $\CM$ between $(\CB,c)$ and $(\CC,c)$. We denote the new gapless edge obtained from this fusion by $(V,\CBs)\boxtimes_{(\CB,c)} \CM$, or graphically as follows:
\be \label{pic:M-sharp}
(V,\CBs)\boxtimes_{(\CB,c)}\CM \quad\quad \xrightarrow{\mbox{represented graphically as}} \quad\quad 
\raisebox{-30pt}{
  \begin{picture}(100,75)
   \put(0,10){\scalebox{0.5}{\includegraphics{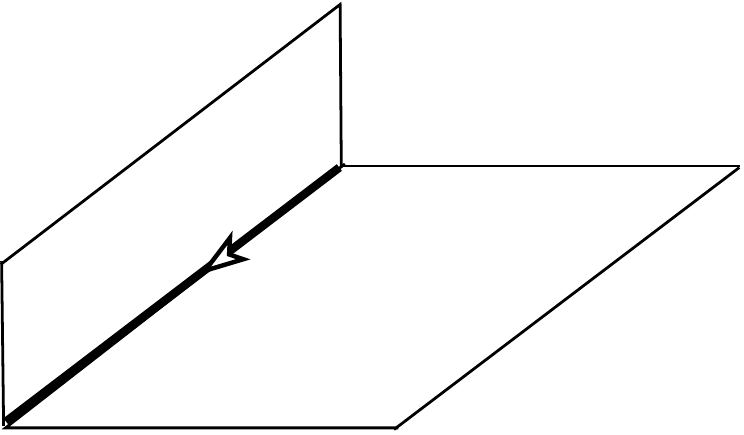}}}
   \put(0,10){
     \setlength{\unitlength}{.75pt}\put(0,0){
     \put(95,40)  {\scriptsize $ (\CC,c) $}
     \put(33,50) {\scriptsize $(\CB,c)$}
     \put(18, 70)  {\scriptsize $(V,\CBs)$}
     \put(38,21)   {\scriptsize $\CM$}
     }\setlength{\unitlength}{1pt}}
  \end{picture}}\, .
\ee
When $\CC=\CB=\CM$, we must have $(V,\CBs) = (V,\CBs)\boxtimes_{(\CB,c)} \CB$. 

\medskip
We will give a detailed analysis of this new edge $(V,\CBs)\boxtimes_{(\CB,c)}\CM$ in \cite{kz3}. We simply state the results here. Recall that the gapped wall $\CM$ between $(\CB,c)$ and $(\CC,c)$ can be described mathematically by a UFC, still denoted by $\CM$, equipped with a unitary braided equivalence $\psi_\CM: \overline{\CB}\boxtimes \CC \to Z(\CM)$ \cite{kk,anyon,ai} (see also \cite{fsv,lww,kawahigashi} for other descriptions). The existence of such a UFC $\CM$ defines an equivalence relation, which is called {\it Witt equivalence}, between the two UMTC's $\CB$ and $\CC$ \cite[Cor.\,5.9]{dmno}. Since the topological edge excitations on the edge $(V,\CBs)$ are labeled by objects in $\CB=\Mod_V$, it is clear that those on the new edge are labeled by objects in $\CB\boxtimes_\CB\CM=\CM$. The tensor unit $\one_\CM=\one_\CB\boxtimes_\CB \one_\CM$ in $\CM$ is the trivial topological edge excitation. Since there is a unitary braided monoidal functor $\phi_\CM: \overline{\CB} \hookrightarrow \overline{\CB}\boxtimes \CC \xrightarrow{\psi_\CM} Z(\CM)$, for $x,y\in \CM$, the internal hom $[x,y]$ is a well-defined object in $\CB$ (see Eq.\,(\ref{eq:internal-hom-1}) or Eq.\,(\ref{diag:univ-prop})). The chiral algebra $U=A_{\one_\CM}$ on this new edge is given by the internal hom $[\one_\CM,\one_\CM]$ in $\CB$, i.e. $U=A_{\one_\CM}=[\one_\CM,\one_\CM]$. It turns out that $[\one_\CM,\one_\CM]$ is a condensable algebra in $\CB$ (see Def.\,\ref{def:sep-alg}). According to \cite{hkl}, $U$ is a VOA extension of $V$. In general, $U\neq V$ (see Remark\,\ref{rem:VM-V}). The boundary CFT $A_x$ is given by the internal hom $[x,x]$ in $\CB$. The defect fields $M_{x,y}$ between the boundary CFT's $[x,x]$ and $[y,y]$ are given by the internal hom $[x,y]$ in $\CB$. As a consequence, the new edge $(V,\CBs)\boxtimes_{(\CB,c)}\CM$ is described by $V$ and a $\CB$-enriched monoidal category $\CMs$, where $\CMs$ is uniquely determined by the pair $(\CB,\CM)$ via the canonical construction given in Example\,\ref{rem:canonical-construction}. Moreover, we proved in \cite[Cor.\,3.3]{kz2} that the Drinfeld center $Z(\CMs)$ of $\CMs$ is exactly $\CC$, i.e. $Z(\CMs)=\CC$. Namely, the boundary-bulk duality still holds for this new gapless edge.

\begin{rem} \label{rem:VM-V}
In general, $U\neq V$. For example, let $A$ be a condensable algebra 
in $\CB$ and $A\neq \one_\CB$. Let $(\CC,c)$ be the 2d bulk phase obtained by condensing $A$ in the 2d bulk phase $(\CB,c)$ \cite{anyon} (see a review \cite{burnell} and references therein), i.e. $\CC=\CB_A^0$ \cite{anyon}, where $\CB_A^0$ is the category of local $A$-modules in $\CB$ \cite{bek,ko,dmno}. This condensation process creates between two phases a gapped domain wall $\CM$ \cite{bs}, described mathematically by the category $\CB_A$ of $A$-modules in $\CB$, i.e. $\CM=\CB_A$ \cite{anyon}. In this case, we have $U=[\one_\CM,\one_\CM]=A\neq \one_\CB=V$. 
\end{rem}

\begin{rem}
On this new edge, all boundary CFT's $[x,x]$ and walls $[x,y]$ preserve only the $V$-symmetry instead of the $U$-symmetry. As a consequence, $\CMs$ is enriched in $\CB=\Mod_V$ instead of in $\Mod_U$ (recall Remark\,\ref{rem:U-non-local}). We will give a detailed explanation of this phenomenon in \cite{kz3}. 
\end{rem}

Since $\CMs$ is uniquely determined by the pair $(\CB, \CM)$, we denote the new edge $(V,\CBs)\boxtimes_{(\CB,c)}\CM$ by the triple $(V,\CB,\CM)$. This notation has a lot of advantages. First, notice that the canonical edge $(V,\CBs)$ of $(\CB,c)$, in the new notation, is just the triple $(V,\CB,\CB)$. Secondly, this notation automatically include the mathematical description of gapped edges as special cases (recall Remark\,\ref{rem:en-in-H}). For example, if $\CN$ is a gapped edge of a 2d bulk phase $(\CD,0)$, it can be expressed as a triple $(\Cb,\bh,\CN)$, where $\Cb$ denotes the field of complex numbers, viewed as the simplest VOA with 0 central charge.  Thirdly, the fusion product $(V,\CBs)\boxtimes_{(\CB,c)}\CM$ can be easily recovered by the following fusion formula
\be \label{eq:fusing-walls-1}
(V,\CBs) \boxtimes_{(\CB,c)}  \CM =  (V,\CB,\CB) \boxtimes_{(\CB,c)} (\Cb,\bh,\CM):= (V\otimes_\Cb \Cb, \CB\boxtimes \bh,\CB\boxtimes_\CB \CM) = (V,\CB,\CM) = (V,\CMs). 
\ee

\begin{figure} 
$$
 \raisebox{-50pt}{
  \begin{picture}(100,100)
   \put(-120,0){\scalebox{0.8}{\includegraphics{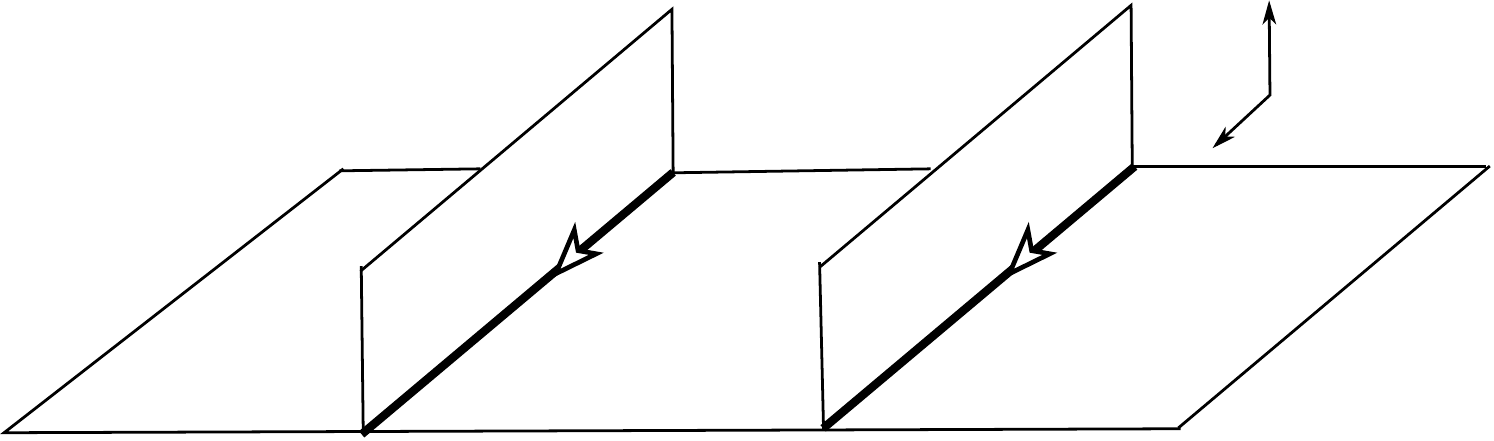}}}
   \put(-120,0){
     \setlength{\unitlength}{.75pt}\put(0,0){
     \put(30,10)  {$ (\CC,c_1) $}
     \put(170,10) {$(\CD,c_1+c_2)$}
     \put(280,10) {$(\CE,c_1+c_2+c_3)$}
     \put(157, 129) {$(U,\CAs)$}
     \put(160, 80)  {$(\CA,c_2)$}
     \put(170,40)   {$\CM$}
     \put(313,40)   {$\CN$}
     \put(297, 129) {$(V,\CBs)$}
     \put(302, 80)  {$(\CB,c_3)$}
     \put(382,129) {$t$}
     \put(367, 95) {$x$}
     }\setlength{\unitlength}{1pt}}
  \end{picture}}
$$
\caption{This picture illustrate a wall $(U,\CA,\CM)$ between $(\CC,c_1)$ and $(\CD,c_1+c_2)$ and a wall $(V,\CB,\CN)$ between $(\CD,c_1+c_2)$ and $(\CE,c_1+c_2+c_3)$. The fusion of these two walls is defined by Eq.\,(\ref{eq:fusing-walls}).  
}
\label{fig:fusing-walls}
\end{figure}

\medskip
Since domain walls can be viewed as special cases of edges by the well-known folding trick, above construction also works for gapless/gapped domain walls. More precisely, for two given bulk phases $(\CC,c_1)$ and $(\CD,c_1+c_2)$ (see Figure\,\ref{fig:fusing-walls}), a gapless domain wall between them can be obtained by fusing a gapped 1d defect $\CM$ at the intersection of $\CC,\CA,\CD$ with the canonical edge $(U,\CAs)$ of the bulk phase $(\CA,c_2)$. This new wall is nothing but $(U,\CAs)\boxtimes_{(\CA,c_2)} \CM = (U, \CA, \CM)$. Now we consider the fusion of two such walls $(U, \CA, \CM)$ and $(V, \CB, \CN)$ depicted in Figure\,\ref{fig:fusing-walls}. This picture suggests that we should fuse those (vertically illustrated) data on these two walls horizontally. This leads to the following fusion formula (which generalizes Eq.\,(\ref{eq:fusing-walls-1})): 
\be \label{eq:fusing-walls}
(U, \CA, \CM) \boxtimes_{(\CD,c_1+c_2)} (V, \CB, \CN) =
(U\otimes_\Cb V, \CA\boxtimes \CB, \CM \boxtimes_\CD \CN),
\ee
where the relative tensor product $\CM \boxtimes_\CD \CN$ is defined in \cite{ta,eno2,bbj1,kz1}. 
We want to remark that $\CM\boxtimes_\CD\CN$ is, in general, not a UFC but a unitary multi-fusion categories even if $\CM$ and $\CN$ are both UFC's \cite{kong-wen-zheng-1,ai}. It describes an unstable 1d gapped defect \cite{kong-wen-zheng-1,ai}. 

\begin{rem} \label{rem:flip}
We have decorated each domain wall in Figure\,\ref{fig:fusing-walls} by an arrow, which represents the orientation of 1+1D world sheets of the wall. More precisely, we introduce a complex coordinate $z=t+ix$ on the 1+1D world sheet as shown in Figure\,\ref{fig:fusing-walls}, where the arrow on the wall coincides with the orientation of the $x$-axis. This arrow also indicates the order of the fusion product of two topological excitations on the wall. For example, if a topological excitation $a$ is located at $(0,0)$ and $b$ located at $(0,1)$ in the $(t,x)$-plane, then we fuse them as $a\otimes b$ (same as the convention in \cite{kk,ai}). If we only flip the arrow, say the one on the $(V,\CB,\CN)$-wall, then the same triple represents a different wall. If we not only flip the arrow but also change the triple $(V,\CB,\CN)$, we can keep the wall physically unchanged. More precisely, in terms of the world sheet coordinate $z=t+ix$, flipping the arrow can be achieved by an orientation-reversing map $t\mapsto t, x\mapsto -x$, or equivalently, $z\mapsto \bar{z}$. We use $\overline{V}$ to denote the same VOA as $V$ but containing only anti-chiral fields $\phi(\bar{z})$ for $\phi\in V$. It is clear that $\CN$ should be replaced by $\CN^\rev$, which is the same category of $\CN$ but equipped with a reversed tensor product $\otimes^\rev$ defined by $a\otimes^\rev b:= b\otimes a$ for $a,b\in \CN$. The braided monoidal functor $\phi_\CN: \overline{\CB} \to Z(\CN)$ is automatically a braided moniodal functor $\phi_\CN: \CB \to \overline{Z(\CN)} = Z(\CN^\rev)$. Also note that changing $V$ to $\overline{V}$ is compatible with the change from $\CB$ to $\overline{\CB}$ according to the construction of braidings in $\CB$ in \cite{huang-mtc2}. Therefore, the triple  $(V,\CB,\CN)$, together with the arrow on the wall, represents the same wall as the triple $(\overline{V}, \overline{\CB}, \CN^\rev)$, together with a flipped arrow. 
\end{rem}

\section{Classification of gapless/gapped edges} \label{sec:classification}

In Section\,\ref{sec:observables}, we have explained that a chiral gapless edge of $(\CC,c)$ should be described by a pair $(V,\CXs)$, where $\CXs$ is an $\Mod_V$-enriched monoidal category for a VOA $V$. Let $\CB=\Mod_V$. The objects in $\CXs$ are topological edge excitations. For $x,y\in \CXs$, the hom space $\Hom_{\CXs}(x,x)$ is a boundary CFT, and $\Hom_{\CXs}(x,y)$ is a wall between two boundary CFT's $\Hom_{\CXs}(x,x)$ and $\Hom_{\CXs}(y,y)$. These boundary CFT's and walls between them all preserve the $V$-symmetry. In general, the chiral algebra $U=\hom_{\CXs}(\one, \one)$, where $\one$ is the trivial topological edge excitation, is a non-trivial extension of $V$ (recall Remark\,\ref{rem:VM-V}). 

\medskip
Mathematically,  it is known, by a nice work \cite{MP}, that any $\CB$-enriched monoidal category $\CXs $ is equivalent to the one obtained from the canonical construction from a pair $(\CB,\CX)$, where $\CX$ is a monoidal category of $\CXs$ equipped with a braided oplax-monoidal functor $\phi_\CX:\overline{\CB} \to Z(\CX)$, i.e. $\CXs=(\CB,\CX)$. We have explained this canonical construction in Example\,\ref{rem:canonical-construction}. In \cite{kz3}, we will argue that the only physical relevant cases are those $(\CB,\CX)$ such that $\CX$ is a UFC and $\phi_\CM$ is a unitary braided monoidal functor. Note that $\phi_\CM$ is automatically fully faithful by \cite[Corollary\,3.26]{dmno}. 

Moreover, $\CXs$ should satisfy the boundary-bulk duality \cite{kong-wen-zheng-2}. Namely, the Drinfeld center $Z(\CXs)$ of $\CXs$ should coincide with the UMTC $\CC$. By \cite[Cor.\,3.3]{kz2}, we have $Z(\CXs)=\CC$ if and only if $Z(\CX)=\overline{\CB}\boxtimes\CC$. This implies that the UFC $\CX$ can be physically realized by a gapped wall between $(\CB,c)$ and $(\CC,c)$. Therefore, the gapless edge $(V,\CXs)$ of $(\CC,c)$ should be nothing but the triple $(V,\CB,\CX)$ constructed in Section\,\ref{sec:general-edges}.

\medskip
Therefore, we propose a classification result of all gapped and chiral gapless stable edges associated to the same 2d bulk phase $(\CC, c)$ as follows: 
\begin{quote}
Gapped and chiral gapless stable edges of $(\CC, c)$ one-to-one correspond to triples $(V,\CB,\CX)$, where $V$ is a VOA $V$ with central charge $c$ such that $\CB=\Mod_V$, and $\CX$ is a unitary fusion category equipped with a unitary braided equivalence $\psi_\CX: \overline{\CB}\boxtimes \CC \xrightarrow{\simeq} Z(\CX)$. The edge is gapped if $V=\Cb$ and gapless if otherwise. Equivalently, one can replace the data $\CX$ by a Lagrangian algebra $A_\CX$ in $\overline{\CB}\boxtimes \CC$. 
\end{quote}
If we allow to include unstable edges, we can simply replace ``fusion" by ``multi-fusion" in the description of $\CX$ (see \cite{kong-wen-zheng-1,ai}). 

\begin{rem} \label{rem:gappable}
When $V=\Cb$, this classification recovers the classification result of gapped edges of $(\CC,0)$. A non-chiral gapless edge of a bulk phase $(\CC,0)$ is gappable if there is UFC $\CX$ such that $Z(\CX)=\CC$. An example of such non-chiral gappable walls is given in Section\,\ref{sec:bcft}. \end{rem}

\begin{rem}
There is a nice classification of gapless edges for abelian 2d topological orders given in \cite{ccmnpy}. It will be very interesting to compare our classification with that in \cite{ccmnpy}. Constructing new gapless edges via the anyon condensations of a non-abelian bulk phase $(\CC,c)$ was considered in some special cases in \cite{bn}. In our constructions of gapless edges, the Witt equivalence relation between $\CC$ and $\CB$ not only includes the cases, in which $(\CB,c)$ is obtained from $(\CC,c)$ via an anyon condensation \cite{anyon}, but also more general ones, in which both $(\CB,c)$ and $(\CC,c)$ are obtained from anyon condensations of another bulk phase $(\CD,c)$ \cite{dmno,anyon}. 
\end{rem}

Our classification proposal reduces the problem of classifying all chiral gapless edges of a given bulk phase $(\CC,c)$ to the problem of classifying all VOA's whose module categories are Witt equivalent to $\CC$. The later problem is still widely open. Interestingly, if every $(\CC,c)$ can indeed be realized by a 2d topological order and if $\CC$ is not Witt equivalent to trivial UMTC $\bk$, then $(\CC,c)$ should have topologically protected chiral gapless edges. It suggests the following mathematical conjecture. 

\begin{conj}
For any UMTC $\CC$, there is a unitary VOA $V$ such that $\Mod_V$ is a UMTC which is Witt equivalent to $\CC$.  
\end{conj}

\begin{rem}
The above conjecture is significantly weaker than the stronger conjecture that every UMTC $\CC$ coincides with $\Mod_V$ for some VOA $V$ such that its central charge mod 8 coincides with the topological central charge of $\CC$ (see for example \cite{tew}). Our theory of chiral gapless edges of 2d topological orders does not require (nor support) the stronger conjecture to be true. Interestingly, 
if there are only finitely many chiral gapless edges of a given bulk phase $(\CC,c)$, an assumption which is not unreasonable, then our theory suggests that there are only finitely many unitary VOA's of a fixed central charge $c$ such that their module categories are Witt equivalent to a given UMTC $\CC$. 
\end{rem}

By the folding trick, we automatically obtain a classification of the gapless/gapped walls between two 2d bulk phases $(\CC,c_1)$ and $(\CD,c_2)$. If $c_1=c_2$ and $\CC,\CD$ are Witt equivalent \cite{dmno,anyon}, then there are gapped walls; if otherwise, then all the walls are necessarily gapless.


\section{0d defects and bulk CFT's} \label{sec:bcft}

In this section, we briefly discuss 0d defects between different gapless edges and the appearance of modular-invariant bulk CFT's. 

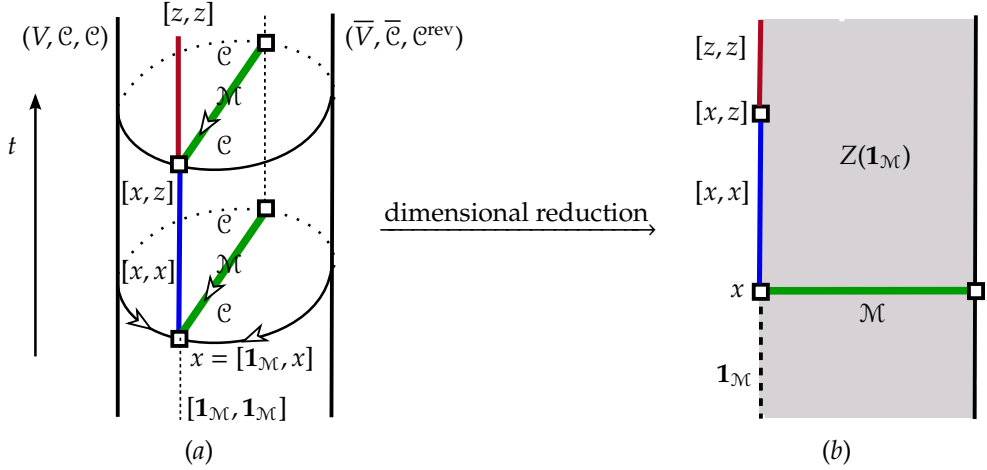
\begin{figure} 
$$
 \raisebox{-70pt}{
  \begin{picture}(100,160)
   \put(0,0){\scalebox{0.6}{\includegraphics{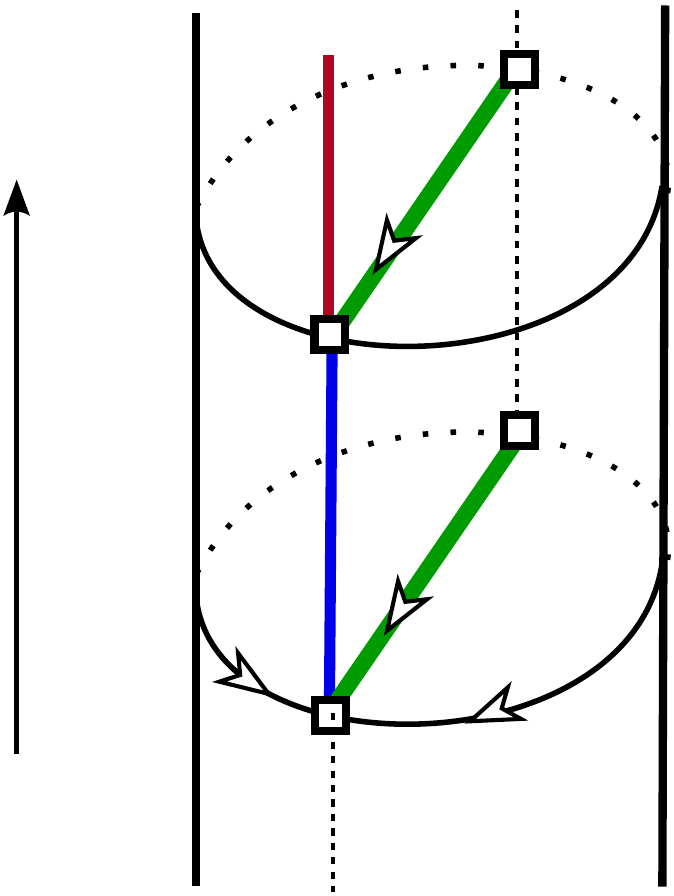}}}
   \put(0,0){
     \setlength{\unitlength}{.75pt}\put(0,0){
     \put(-10, 135) {$t$}
     \put(68,202)  {$[z,z]$}
     \put(47,112)  {$[x,z]$}
     \put(47,72)  {$[x,x]$}
     \put(80,3)    {$[\one_\CM,\one_\CM]$}
     \put(95,180)  {$\CC$}
     \put(95,135)  {$\CC$}
     \put(95, 160) {$\CM$}
     \put(95, 95)  {$\CC$}
     \put(95,50)  {$\CC$}
     \put(95, 75) {$\CM$}
     \put(-3,190) {$(V,\CC,\CC)$}
     \put(160,190) {$(\overline{V}, \overline{\CC},\CC^\rev)$}
     \put(82,28)  {$x=[\one_\CM,x]$}
     }\setlength{\unitlength}{1pt}}
  \end{picture}}
\quad\quad\quad \xrightarrow{\mbox{dimensional reduction}} \quad\quad\quad
 \raisebox{-70pt}{
  \begin{picture}(100,160)
   \put(0,0){\scalebox{0.6}{\includegraphics{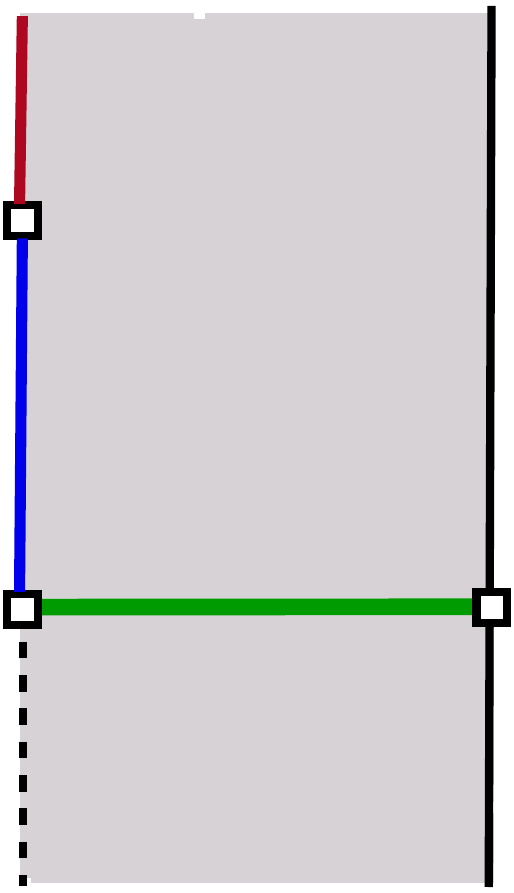}}}
   \put(0,0){
     \setlength{\unitlength}{.75pt}\put(-40,0){
     \put(12,152)  {$[x,z]$}
     \put(12,112)  {$[x,x]$}
     \put(12, 185) {$[z,z]$}
     \put(30,63)  {$x$}
     \put(95,50)  {$\CM$}
     \put(23,20) {$\one_\CM$}
     \put(85,130) {$Z(\one_\CM)$}
     }\setlength{\unitlength}{1pt}}
  \end{picture}}  
$$
$$
(a) \quad\quad\quad\quad \quad \quad\quad\quad\quad \quad\quad \quad\quad\quad\quad\quad
\quad\quad\quad
\quad\quad\quad\quad  (b)
$$
\caption{ (a) depicts a two 2+1D bulk phases $(\CC,c)$ and $(\CC,c)$, equipped with the canonical edge and its time reverse, respectively, and separated by a gapped domain wall $\CM$. $x,z$ are objects in $\CM$, and the internal hom $[x,z] \in \CM$ is defined by $[x,z]=z\otimes x^\ast$. These internal homs form a canonical 0d edge of the 1d wall $\CM$ (more details will be given in \cite{kz3}). (b) depicts a 1+1D world sheet obtained by squeezing the entire 2-disk in (a) to the 1d wall $\CM$. 
}
\label{fig:cl-CFT}
\end{figure}

\medskip
Let us consider the situation depicted in Figure\,\ref{fig:cl-CFT} (a). Let $\CM$ be a gapped wall between two bulk phases $(\CC,c)$ and $(\CC,c)$. Namely, $\CM$ is a UFC such that $Z(\CM)=\CC\boxtimes \overline{\CC}=Z(\CC)$ as UMTC's. Let $(V, \CC,\CC)$ be the canonical edge of $(\CC,c)$ (i.e. $\Mod_V=\CC$). Note that we have flipped the orientation of the part of edge on the right side of the wall $\CM$. As we have discussed in Remark\,\ref{rem:flip}, we need change the triple $(V, \CC,\CC)$ to $(\overline{V},\overline{\CC},\CC^\rev)$ in order to leave the edge unchanged. 

Now we consider a dimensional reduction process by fusing these two gapless edges with the gapped wall $\CM$. The process is depicted as passing from Figure\,\ref{fig:cl-CFT} (a) to (b). The result of this fusing process is a 1d gapless wall between two trivial phases defined by
\be \label{eq:C-M-C}
(V, \CC,\CC) \boxtimes_{(\CC,c)} (\Cb,\bh,\CM) \boxtimes_{(\CC,c)} (\overline{V}, \overline{\CC}, \CC^\rev)
= (V\otimes_\Cb \overline{V}, \CC\boxtimes \overline{\CC}, \CM). 
\ee
Its 1+1D world sheet is depicted in Figure\,\ref{fig:cl-CFT} (b). Both chiral and anti-chiral fields live on this 1+1D world sheet. Those fields on the world line supported on the trivial topological excitation $\one_\CM$ are given by the internal hom $[\one_\CM,\one_\CM]_{\CC\boxtimes \overline{\CC}}$ in $\CC\boxtimes \overline{\CC}$. This internal hom is a Lagrangian algebra and is also called the full center of $\one_\CM$, denoted by $Z(\one_\CM)$. It is not a VOA but a modular-invariant bulk CFT over $V\otimes_\Cb \overline{V}$ \cite{fjfrs,kr2}. When $\CM=\CC$, $Z(\one_\CC)=\oplus_{i\in I(\CC)} \, i\boxtimes i^\ast$, where $I(\CC)$ is the set of simple objects in $\CC$, is nothing but the charge-conjugate modular invariant bulk CFT. It turns out that the map $\CM \mapsto Z(\one_\CM)$ defines a one-to-one correspondence between gapped walls $\CM$ and modular-invariant bulk CFT's over $V\otimes_\Cb \overline{V}$. Therefore, Figure\,\ref{fig:cl-CFT} provides a physical explanation of the one-to-one correspondences among the following three sets: the set of Lagrangian algebras in $Z(\CC)$, that of gapped walls between $(\CC,c)$  and $(\CC,c)$, and that of modular-invariant bulk CFT's in $\CC\boxtimes \overline{\CC}$ \cite{kr1,kr2,dmno,anyon}.

\begin{rem}
More general heterotic modular invariant bulk CFT's will be discussed in Remark\,\ref{rem:C-M-D}. The correspondence between Lagrangian algebras and modular-invariant bulk CFT's was first proved in \cite[Thm.\,3.4]{kr2}. The physical explanation of this correspondence was provided for abelian anyon systems by Levin in \cite{levin}. 
Using the general anyon condensation theory \cite{anyon} (see also \cite{erb,hw}), Levin's arguments can be trivially generalized to non-abelian anyon systems. In this work, we have generalized all these earlier works, and have made this correspondence more precise by an explicit formula Eq.\,(\ref{eq:C-M-C}) and by its generalization in Remark\,\ref{rem:C-M-D} in terms of a new mathematical structure: enriched monoidal category. 
\end{rem}

Now we discuss a 0d edge of the 1d wall $\CM$. In order to give a boundary CFT interpretation of the observables on this 0d edge, we must link $\CM$ to an OSVOA. It turns out that $\CM$ can always be realized by the category $\CC_{W|W}$ of $W$-$W$-bimodules, where $W$ is a CSSFA in $\CC$ \cite{ostrik,eno1,fs1,frs1} thus an OSVOA extension of $V$ \cite{osvoa}. As a consequence, $\CM$ can be viewed as a subcategory of $\CC$. This CSSFA $W$ is not unique, but it is unique up to Morita equivalence. Different choices of $W$ in the same Morita class amount to relabeling the objects in $\CM$. Moreover, there is a one-to-one correspondence between the set of UFC's $\CM$ (up to equivalences) such that $Z(\CM)=Z(\CC)$ and that of the Morita classes of CSSFA's in $\CC$ \cite{eno1}.

Once we have fixed $W$, there is a canonical 0d edge of the gapped wall $\CM$ (depicted in Figure\,\ref{fig:cl-CFT} (a)) defined by a pair $(W,\CMs)$, where $\CMs$ is an $\CM$-enriched category. More precisely, the objects in $\CMs$ are the same as those in the UFC $\CM$, and the hom spaces $\Hom_{\CMs}(x,z)$ are given by the internal hom $[x,z]=z\otimes x^\ast$ in $\CM$ (recall Example\,\ref{exam:can-const-0}). The internal homs $[x,x]$ for $x\in\CM$ are CSSFA's in $\CM$, and automatically OSVOA extensions of $W$. They are automatically boundary CFT's over $V$ \cite{ffrs,kr1,kr2}. The internal homs $[x,z]$ are walls between boundary CFT's. By passing from Figure\,\ref{fig:cl-CFT} (a) to (b), this 0d edge $(W,\CMs)$ becomes a 0d edge of the 1d gapless wall $(V\otimes_\Cb \overline{V}, \CC\boxtimes \overline{\CC}, \CM)$ between two trivial phases. Its 0+1 world line is depicted as the left boundary of the 1+1D world sheet in Figure\,\ref{fig:cl-CFT} (b). Boundary CFT's on this world line are CSSFA's $\one_\CM, [x,x], [z,z]$ in $\CM$, and they must share the same bulk CFT $Z(\one_\CM)$. Mathematically, it means that they must have the same full center, i.e. $Z([x,x])=Z([z,z])=Z(\one_\CM)$. It is guaranteed because they are all Morita equivalent \cite{ffrs,kr1,davydov}. Moreover, these internal homs $[x,x]$ in $\CM$ for all $x\in \CM$ realize all boundary CFT's over $V$ that share the same bulk CFT $Z(\one_\CM)$ \cite{kr1,kr2}. 

\begin{rem} \label{rem:cl-cft} 
In particular, when $\CM=\CC$, $[x,x]=x\otimes x^\ast$ for $x\in \CC$ are all the CSSFA's in $\CC$ that are allowed on the canonical edge $(V,\CCs)$ (recall Remark\,\ref{rem:xx-zz}). When $\CM\neq \CC$, $[x,x]\in \CM=\CC_{W|W}$ recovers CSSFA's in $\CC$ in a Morita class different from that of $\one_\CC$. 
\end{rem}

In this way, we have naturally recovered all boundary-bulk CFT's over $V$ from 2d topological orders via a dimensional reduction process.

\begin{rem}
Since $(V\otimes_\Cb \overline{V}, \CC\boxtimes \overline{\CC}, \CM)$ is a gapless wall between two trivial phases, the Drinfeld center of the enriched monoidal category $(\CC\boxtimes \overline{\CC},\CM)$ is exactly the trivial UMTC $\bh$ \cite{kz2}. Such a gapless wall is called gappable (recall Remark\,\ref{rem:gappable}). By \cite[Corollary\,4.3]{zheng}, the enriched monoidal category $(Z(\CM),\CM)$ is Morita equivalent to $(\bh,\bh)$, and the Morita equivalence is exactly defined by the invertible $(\bh,\bh)$-$(Z(\CM),\CM)$-bimodule $\CMs$. 
\end{rem}

\begin{rem}  \label{rem:C-M-D}
More generally, bulk phases on the two sides of the gapped wall $\CM$ in Figure\,\ref{fig:cl-CFT} (a) can be different, say $(\CC,c)$ and $(\CD,c)$. There is a VOA $A$ with two VOA extensions $U$ and $V$ such that $\Mod_U=\CC$, $\Mod_V=\CD$ \cite{dmno}. The internal hom $[\one_\CM,\one_\CM]_{\CC\boxtimes\overline{\CD}}=Z(\one_\CM)$ is a Lagrangian algebra in $\CC\boxtimes\overline{\CD}$. The map $\CM\mapsto Z(\one_\CM)$ defines a bijection between the set of gapped walls between $(\CC,c)$ and $(\CD,c)$, that of Lagrangian algebras in $\CC\boxtimes\overline{\CD}$. These Lagrangian algebras are precisely those modular-invariant bulk CFT's in $Z(\Mod_A)$ containing $U\otimes_\Cb \overline{V}$ as a full field subalgebra \cite{ffrs2,davydov2,dmno}. The fusion category $\CM$ can be realized by $\CE_{W|W}$,  where $W$ is a CSSFA in $\CE:=(\Mod_A)_{U|V}$. The 0d edge of the wall $\CM$ is still given by the pair $(W,\CMs)$, where $\CMs$ is the $\CM$-enriched category consisting of the same objects as those in $\CM$ and $\Hom_{\CMs}(x,z):=[x,z]=z\otimes x^\ast\in\CM$. All $[x,z]$ in $\CM$ can be interpreted as boundary CFT's (or walls) over $A$ and share the same bulk CFT $Z(\one_\CM)$. We will give more details in \cite{kz3}. 
\end{rem}

\begin{rem} 
We will give a detailed study and a classification of all 0d walls in \cite{kz3}. There are some physics works on 0d walls between different gapless edges of 2d topological orders (see \cite{ccbcn}). But boundary CFT's were not mentioned there. 
\end{rem}

\begin{rem}
It will be interesting to understand the relation between our approach towards the gapless edges of 2d topological orders and the well-known formulation of boundary-bulk RCFT's as holographic boundaries of a 2+1D Reshetikhin-Turaev TQFT's \cite{fffs,fs1,frs1,frs2,ffrs2,ks}. We hope to clarify this relation in the future. 
\end{rem}

\section{Summary and outlooks} \label{sec:summary}

In this work, we have shown that a gapless edge of a 2d bulk phase $(\CC,c)$ can be mathematically described/classified by an enriched monoidal category $\CMs$ together with a VOA $V$. This description includes that of gapped edges as special cases. Therefore, we have found a unified mathematical theory for both gapped and gapless edges.

We have left out a few important issues that will be discussed in \cite{kz3}. We briefly mention them below. 
\begin{itemize}
\item Some gapless edges given in our classification are actually gappable. A gapless edge is gappable if it shares the same bulk with a gapped edge. Mathematically, a gapless edge $(V,\CB,\CM)$ is gappable if its Drinfeld centers $Z(\CB,\CM)$ coincide with that of a UFC $\CN$, or equivalently, $(\CB,\CM)$ is Morita equivalent to $(\bh,\CN)$ \cite{zheng}.

\item In \cite{kz1}, the complete mathematical statement of the boundary-bulk duality for gapped edges can be stated as a functor mapping UFC's to their Drinfeld centers. Moreover, the functor is fully faithful. We will prove a similar result for gapless/gapped edges. 

\end{itemize}

Enriched monoidal categories are also useful in the study of topological order with symmetries (SPT's/SET's). Our results shed light on how to describe gapless/gapped edges for SPT's and SET's. We will come back to this point in the future.  

The unified mathematical theory of gapless/gapped edges/walls also allows us to compute global observables for topological orders with gapless edges/walls via factorization homology \cite{ai}. It will be interesting to explore them thoroughly in the future. 

It is very important to study how to obtain a gapless edge from another via pure edge phase transitions (see for example \cite{pmn,ccbcn,ccmnpy} and references therein). 


\appendix

\section{Appendix}  

\subsection{Algebras in unitary modular tensor categories}  \label{sec:alg-umtc}
In this subsection, we recall the definitions of a few types of algebras in a UFC or a UMTC. 

\medskip
A unitary fusion category (UFC) $\CC$ is a unitary finite abelian rigid monoidal category \cite{eno0, egno}.  We will not recall its definition, but only recall a few important ingredients of it. 
\bnu
\item there are finitely many simple objects, and all objects are direct sums of simple objects; 
\item the hom spaces $\hom_\CC(x,y)$ are all finite dimensional Hilbert spaces; For every morphism $f: x\to y$, there is an adjoint $f^\ast: y\to x$;  and $f^\ast \circ f =0$ implies that $f=0$; 
\item there is an associative tensor product $\otimes: \CC \times \CC \to \CC$, i.e. $x\otimes (y\otimes z) \simeq (x\otimes y) \otimes z$, and a distinguished object $\one$ called the tensor unit; $\one$ is simple;  
\item each object $x$ has a (two-side) dual object $x^\ast$, together with duality maps $u_x: \one \to x\otimes x^\ast$ and $v_x: x^\ast \otimes x \to \one$. 
\enu
A UFC $\CC$ is a unitary braided fusion category if it is also equipped with a braiding $c_{x,y}: x\otimes y \xrightarrow{\simeq} y\otimes x$ for $x,y\in \CC$ satisfying the Hexagon relations. A UMTC is a unitary braided fusion category such that $\one$ is the only simple object that is symmetric to all objects \cite{turaev}. In other words, if $x$ is simple and $c_{y,x}\circ c_{x,y} = \Id_{x\otimes y}, \forall y\in \CC$, then $x\simeq \one$.

\begin{defn}  \label{def:alg}
Let $\CC$ be a monoidal category. An {\it algebra} in $\CC$ is a triple $(A,m,\iota)$, where $A$ is an object in $\CC$, $m: A\otimes A \to A$ and $\iota: \one\to A$ are morphisms in $\CC$, such that 
$$
m\circ (m\otimes \Id_A) = m\circ (\Id_A \otimes m), \quad\quad  m\circ (\iota \otimes \Id_A) = \Id_A = m \circ (\Id_A \otimes \iota). 
$$
If $\CC$ is also braided, then $A$ is called {\it commutative} if $m\circ c_{A,A} =m$. 
\end{defn}

\begin{defn} \label{def:sep-alg}
An algebra $(A,m,\iota)$ in a monoidal category $\CC$ is called {\it separable} if there is an $A$-$A$-bimodule map $e: A\to A\otimes A$ such that $m\circ e = \Id_A$. A separable algebra $(A,m,\iota)$ is called {\it connected} if $\hom_\CC(\one,A) = \Cb$. If $\CC$ is a UMTC, then a commutative connected separable algebra is called an {\it \'{e}tale algebra} \cite{dmno}, or a {\it condensable algebra} (due to the fact that it determines an anyon condensation in a 2d topological order \cite{anyon}). 
\end{defn}

\begin{defn} [\cite{dmno}] \label{def:lag-alg}
A {\it Lagrangian algebra} in a UMTC $\CC$ is a condesable algebra $A$ such that $(\dim A)^2 = \dim (\CC)$. 
\end{defn}

\begin{defn}
Let $\CC$ be a monoidal category. A co-algebra in $\CC$ is a triple $(A,\Delta,\epsilon)$, where $A$ is an object in $\CC$, $\Delta: A\to A\otimes A$ and $\epsilon: A \to \one$ are morphisms in $\CC$, such that 
$$
(\Delta \otimes \Id_A) \circ \Delta = (\Id_A \otimes \Delta) \circ \Delta, \quad\quad 
(\epsilon \otimes \Id_A) \circ \Delta = \Id_A = (\Id_A \otimes \epsilon) \circ \Delta. 
$$
\end{defn}

\begin{rem}
If $\CC$ is UFC and $(A,m,\iota)$ an algebra in $\CC$, then the triple $(A,m^\ast, \iota^\ast)$ is automatically a co-algebra in $\CC$. 
\end{rem}

\begin{defn}[\cite{fs1}] \label{def:ssfa}
Let $\CC$ be a monoidal category. A {\it Frobenius algebra} in $\CC$ is a quintuple $(A,m,\iota,\Delta, \epsilon)$ such that $(A,m,\iota)$ is an algebra and $(A,\Delta,\epsilon)$ is a co-algebra and 
$$
(m\otimes \Id_A) \circ (\Id_A \otimes \Delta) = \Delta \circ m = (\Id_A\otimes m) \circ (\Delta \otimes \Id_A). 
$$
Such a Frobenius algebra is called {\it special} if $m\circ \Delta = \Id_A$. When $\CC$ is a UFC, a Frobenius algebra in $\CC$ is called {\it symmetric} if 
$$
(\Id_{A^\ast} \otimes (\epsilon\circ m)) \circ (u_{A^\ast} \otimes \Id_A)  = ((\epsilon \circ m)\otimes \Id_{A^\ast}) \circ (\Id_A \otimes u_A). 
$$
We abbreviate a {\it connected special symmetric Frobenius algebra} to a CSSFA. 
\end{defn}

\begin{rem} \label{rem:cssfa}
A special Frobenius algebra is automatically a separable algebra. Conversely, a connected separable algebra in a UFC has a unique structure of a CSSFA. 
\end{rem}

Let $\CC$ be a UFC. Let $\CM$ be a left $\CC$-module, which is a unitary category $\CM$ equipped with a unital and associative action $\odot: \CC \times \CM \to \CM$. In other words, for $a, b\in \CC, x\in \CM$, we have $b\odot x\in \CM$, $\one \odot x \simeq x$ and $a\odot (b \odot x) \simeq (a\otimes b) \odot x$. For $x,y\in \CM$, the internal hom $[x,y]$ is an object in $\CC$ defined by the following isomorphisms for all $a\in \CC$,  
\be \label{eq:internal-hom-1}
\gamma_{a,x,y}: \hom_\CM(a\odot x, y) \xrightarrow{\simeq} \hom_\CC(a, [x,y]), 
\ee
which are natural with respect to all three variables $a,x,y$. Equivalently, one can define the internal hom $[x,y]$ by its universal property. When $a=[x,y]$, $\rho:=\gamma_{[x,y],x,y}^{-1}(\Id_{[x,y]})$ is a morphism $[x,y]\odot x \to y$. The pair $([x,y], \rho)$ satisfies the following universal property. If $(u, f)$ is another such a pair, i.e. $f: u\odot x \to y$, then there is a unique morphism $f': u \to [x,y]$ in $\CC$ such that the following diagram 
\be \label{diag:univ-prop}
\xymatrix@R=1.5em{
& [x,y] \odot x \ar[rd]^{\rho} &  \\
u\odot x \ar[rr]^{f} \ar[ru]^{f'\odot \Id_x} & & y
}
\ee
is commutative. This universal property of internal hom determines the pair $([x,y], \rho)$ up to isomorphisms. This universal property also provides a canonical morphism $\ev: [y,z] \otimes [x,y] \to [x,z]$ induced from the action $([y,z]\otimes [x,y])\odot x \to [y,z]\odot y \to z$, and a canonical morphism $\one \to [x,x]$ induced from the unital action $\one \odot x \simeq x $. These morphisms provide $[x,x]$ with a structure of an algebra in $\CC$ and $[x,y]$ with a structure of a $[y,y]$-$[x,x]$-bimodule.

\begin{exam}
When $\CM=\CC$ is viewed as a left $\CC$-module, then $[x,y] = y\otimes x^\ast$ and 
$$
\rho: [x,y]\odot x= y\otimes x^\ast \otimes x \xrightarrow{\Id_y\otimes v_x} x,
$$ 
In this case, it is very easy to show that such defined $[x,y]$ satisfies the universal property of internal hom. More precisely, by the rigidity of $\CC$, there is a canonical isomorphism $\phi: \hom_\CC(u\otimes x, y) \simeq \hom_\CC(u, y\otimes x^\ast)$. For any $f: u\otimes x \to y$, we obtain $f'=\phi(f): u \to y\otimes x^\ast$ such that diagram in Eq.\,(\ref{diag:univ-prop}). Moreover, in this case, the canonical morphism $\ev: [y,z] \otimes [x,y] \to [x,z]$ is explicitly defined by 
$$
z\otimes y^\ast \otimes y \otimes x^\ast \xrightarrow{\Id_z\otimes v_y \Id_{x^\ast}} z \otimes x^\ast,  
$$
and the $\one \to [x,x]$ is defined by $u_x: \one \to x\otimes x^\ast$. It turns out that $[x,x]$ for $x\in \CM$ are CSSFA's in $\CM$. 
\end{exam}

\begin{exam}  \label{rem:full-center}
Let $\CM$ be a UFC. Its Drinfeld center $Z(\CM)$ is a UMTC. $\CM$ is also a left $Z(\CM)$-module. The internal hom $[\one_\CM, \one_\CM]$ in $Z(\CM)$ is a condensable algebra (see Def.\,\ref{def:sep-alg}), and is also called the {\it full center} of $\one_\CM$, denoted by $Z(\one_\CM)$. When $\CM$ is a UMTC, $Z(\CM)\simeq \CM\boxtimes\overline{\CM}$, and $Z(\one_\CM) = \oplus_i i\boxtimes i^\ast$, where $i$ are simple objects in $\CC$, is a Lagrangian algebra in $Z(\CM)$. More generally, the notion of the full center $Z(A)$ can be defined for any algebra $A$ in $\CM$ \cite{davydov}. If $A$ is a CSSFA in $\CM$, then the full center $Z(A)$ of $A$ in $Z(\CM)$ is a Lagrangian algebra in $Z(\CM)$ \cite{fjfrs,kr1,kr2}. 
\end{exam}

\subsection{Enriched monoidal categories and a canonical construction}
In this subsection, we recall the definition of an enriched monoidal category and a canonical construction of it. 

\begin{defn} \label{def:en-cat}
Let $\CB$ be a monoidal category. A {\em category $\CC^\sharp$ enriched in $\CB$}, or a {\it $\CB$-enriched category}, consists of a set of objects $Ob(\CC^\sharp)$, an object $\Hom_{\CC^\sharp}(x,y)$ in $\CB$ for every pair $x,y\in\CC^\sharp$, and a composition morphism $\circ:\Hom_{\CC^\sharp}(y,z)\otimes\Hom_{\CC^\sharp}(x,y)\to\Hom_{\CC^\sharp}(x,z)$ for $x,y,z\in\CC^\sharp$, such that there exists a morphism $\id_x:\one_\CB\to\Hom_{\CC^\sharp}(x,x)$ for $x\in\CC^\sharp$ rendering the following diagrams commutative for $x,y,z,w\in\CC^\sharp$:
\be \label{diag:right-unit}
\xymatrix@!C=15ex{
  & \Hom_{\CC^\sharp}(x,y)\otimes\Hom_{\CC^\sharp}(x,x) \ar[rd]^\circ \\
  \Hom_{\CC^\sharp}(x,y) \ar[rr]^\Id \ar[ru]^{\Id\otimes\id_x} && \Hom_{\CC^\sharp}(x,y), \\
}
\ee
\be \label{diag:left-unit}
\xymatrix@!C=15ex{
  & \Hom_{\CC^\sharp}(y,y)\otimes\Hom_{\CC^\sharp}(x,y) \ar[rd]^\circ \\
  \Hom_{\CC^\sharp}(x,y) \ar[rr]^\Id \ar[ru]^{\id_y\otimes\Id} && \Hom_{\CC^\sharp}(x,y), \\
}
\ee
\be \label{ax:asso-circ}
\xymatrix{
  \Hom_{\CC^\sharp}(z,w)\otimes\Hom_{\CC^\sharp}(y,z)\otimes\Hom_{\CC^\sharp}(x,y) \ar[r]^-{\Id\otimes\circ} \ar[d]_{\circ\otimes\Id} & \Hom_{\CC^\sharp}(z,w)\otimes\Hom_{\CC^\sharp}(x,z) \ar[d]^\circ \\
  \Hom_{\CC^\sharp}(y,w)\otimes\Hom_{\CC^\sharp}(x,y) \ar[r]^-\circ & \Hom_{\CC^\sharp}(x,w). \\
}
\ee
\end{defn}

\begin{rem}
In this work, we distinguish two notations $\Id_x$ and $\id_x$, where $\Id_x$ is the usual identity morphism $x\to x$ in an ordinary category, but $\id_x$ is reserved for the morphism $\one_\CB \to \hom_\CCs(x,x)$ in $\CB$ for a $\CB$-enriched category $\CCs$. 
\end{rem}

\begin{exam} \label{exam:can-const-0} ({\bf Canonical construction I}): 
Let $\CB$ be a UFC and $\CM$ a left $\CB$-module. The categorical structure $\CMs$ consisting of 
\bnu
\item objects in $\CMs$ are objects in $\CM$, i.e. $Ob(\CMs)=Ob(\CM)$; 
\item $\hom_\CMs(x,y)=[x,y]\in \CB$; 
\item $\id_x: \one_\CB \to [x,x]$ is the morphism canonically induced from the unital action $\one_\CB\odot x \simeq x$; 
\item $\circ: [y,z] \otimes [x,y] \to [x,z]$ is the morphism $\ev$ canonically induced from the action $([y,z]\otimes [x,y])\odot x \to [y,z]\odot y \to z$. 
\enu
It is well-known that this $\CMs$ is a $\CB$-enriched category \cite{kelly}. 
\end{exam}

\begin{defn} \label{def:underlying}
The {\it underlying category} $\CC$ of a $\CB$-enriched category $\CCs$ is an ordinary category defined as follows: $Ob(\CC)=Ob(\CCs)$ and $\hom_\CC(x,y):=\hom_\CB(\one_\CB, \hom_\CCs(x,y))$ for $x,y\in Ob(\CC)$. The identity morphism in $\hom_\CC(x,x)$ is $\id_x$ and the composition of morphisms is naturally induced from that of $\CCs$.
\end{defn}

\begin{defn}
An {\em enriched functor} $F:\CC^\sharp\to\CD^\sharp$ between $\CB$-enriched categories consists of a map $F:Ob(\CC^\sharp)\to Ob(\CD^\sharp)$ and a morphism $F:\Hom_{\CC^\sharp}(x,y)\to\Hom_{\CD^\sharp}(F(x),F(y))$ for every pair $x,y\in\CC^\sharp$ such that the following diagrams commute for $x,y,z\in\CC^\sharp$
$$
\xymatrix{
  & \one_\CB \ar[ld]_{\id_x} \ar[rd]^{\id_{F(x)}} \\
  \Hom_{\CC^\sharp}(x,x) \ar[rr]^-F && \Hom_{\CD^\sharp}(F(x),F(x)), \\
}
$$
\be \label{diag:fun-composition}
\xymatrix{
  \Hom_{\CC^\sharp}(y,z)\otimes\Hom_{\CC^\sharp}(x,y) \ar[r]^-\circ \ar[d]_{F\otimes F} & \Hom_{\CC^\sharp}(x,z) \ar[d]^F \\
  \Hom_{\CD^\sharp}(F(y),F(z))\otimes\Hom_{\CD^\sharp}(F(x),F(y)) \ar[r]^-\circ & \Hom_{\CD^\sharp}(F(x),F(z)). \\
}
\ee
\end{defn}
It is clear that the composition of two enriched functors is again an enriched functor. The enriched functor $F: \CCs \to \CDs$ naturally induces an ordinary functor $F: \CC \to \CD$ between two underlying categories.

\begin{defn}
An {\em enriched natural transformation} $\xi:F\to G$ between two enriched functors $F,G:\CC^\sharp\to\CD^\sharp$ consists of a morphism $\xi_x: \one_\CB \to \hom_{\CD^\sharp}(F(x),G(x))$ for $x\in \CC$ such that the following diagram commutes for $x,y\in\CC^\sharp$:
\be \label{eq:en-natural-transformation}
\xymatrix@!C=30ex{
  \Hom_{\CC^\sharp}(x,y) \ar[r]^-{G} \ar[d]_{F} & \Hom_{\CD^\sharp}(G(x),G(y)) \ar[d]^{-\circ\xi_x} \\
  \Hom_{\CD^\sharp}(F(x),F(y)) \ar[r]^-{\xi_y\circ-} & \Hom_{\CD^\sharp}(F(x),G(y)). \\
}
\ee
\end{defn}
Note that the composition of two enriched natural transformations $\xi: F\to G$ and $\eta: G \to H$ is defined by $(\eta\circ \xi)_x=\eta_x \circ \xi_x: \one_\CB \to \Hom_{\CD^\sharp}(F(x), H(x))$. An enriched natural transformation $\xi$ is called an enriched natural isomorphism if each $\xi_x$ is an isomorphism. 

\medskip
Now we assume that $\CB$ is a braided monoidal category equipped with braiding $c_{x,y}:x\otimes y\to y\otimes x$ for $x,y\in \CB$. Let $\CC^\sharp$ and $\CD^\sharp$ be $\CB$-enriched categories. The {\em Cartesian product} $\CC^\sharp\times\CD^\sharp$  is a $\CB$-enriched category defined as follows: 
\begin{itemize}
  \item $Ob(\CC^\sharp\times\CD^\sharp)=Ob(\CC^\sharp)\times Ob(\CD^\sharp)$;
  \item $\Hom_{\CC^\sharp\times\CD^\sharp}((x,y),(x',y')) = \Hom_{\CC^\sharp}(x,x')\otimes\Hom_{\CD^\sharp}(y,y')$;
  \item the composition
\begin{align*}
\circ: \Hom_{\CC^\sharp\times\CD^\sharp}((x',y'),(x'',y'')) \otimes \Hom_{\CC^\sharp\times\CD^\sharp}&((x,y),(x',y')) \\
\to & \Hom_{\CC^\sharp\times\CD^\sharp}((x,y),(x'',y''))
\end{align*} 
is given by 
\begin{align}
&\Hom_{\CC^\sharp}(x',x'')\otimes\Hom_{\CD^\sharp}(y',y'')\otimes\Hom_{\CC^\sharp}(x,x')\otimes\Hom_{\CD^\sharp}(y,y') \nn
&\hspace{1cm} \xrightarrow{\Id\otimes c^{-1}\otimes\Id}  \Hom_{\CC^\sharp}(x',x'')\otimes\Hom_{\CC^\sharp}(x,x')\otimes\Hom_{\CD^\sharp}(y',y'')\otimes\Hom_{\CD^\sharp}(y,y') \\
&\hspace{1cm} \xrightarrow{~~\circ\otimes\circ~~}  \Hom_{\CC^\sharp}(x,x'')\otimes\Hom_{\CD^\sharp}(y,y''). \nonumber
\end{align} 
\end{itemize}

\begin{defn} \label{def:emc}
A {\em $\CB$-enriched monoidal category} consists of a $\CB$-enriched category $\CC^\sharp$, a distinguished object $\one_{\CC^\sharp}$, an enriched functor $\otimes:\CC^\sharp\times\CC^\sharp\to \CC^\sharp$, and enriched isomorphisms $\lambda:\one_{\CC^\sharp}\otimes- \to \Id_{\CC^\sharp}$, $\rho: -\otimes\one_{\CC^\sharp} \to \Id_{\CC^\sharp}$, $\alpha: \otimes\circ(\otimes\times\Id_{\CC^\sharp}) \to \otimes\circ(\Id_{\CC^\sharp}\times\otimes)$ such that the underlying category $\CC$, together with $\otimes$, $\lambda, \rho, \alpha$, defines a monoidal category. 
\end{defn}

\begin{exam} [{\bf Canonical construction II}] \label{rem:canonical-construction}
Let $\CM$ be a monoidal category and $\CB$ a braided monoidal such that there is a braided oplax-monoidal functor $\phi_\CM: \overline{\CB} \to Z(\CM)$, where $Z(\CM)$ is the Drinfeld center of $\CM$. The oplax-monoidal structure of $\phi_\CM$ is a morphism $\beta_{a,b}: \phi_\CM(a\otimes b) \to \phi_\CM(a) \otimes \phi_\CM(b)$ and an isomorphism $\phi_\CM(\one_\CB)\simeq \one_{Z(\CM)}$ satisfying some natural conditions. If $\beta_{a,b}$ for $a,b\in \CB$ are isomorphisms, then an oplax-monoidal functor becomes a monoidal functor. In general, $\beta_{a,b}$ are not isomorphisms. There is a functor $\odot: \overline{\CB} \times \CM \to \CM$ defined by 
$\overline{\CB} \times \CM \xrightarrow{\phi_\CM \times \Id_\CM} Z(\CM) \times \CM \to \CM \times \CM \xrightarrow{\otimes} \CM$.
There is a canonical construction of a $\CB$-enriched monoidal category $\CMs$ from the pair $(\CC,\CM)$ \cite{MP}: 
\begin{itemize}
\item objects in $\CMs$ are objects in $\CM$, i.e. $Ob(\CMs):=Ob(\CM)$; 
\item For $x,y\in \CM$, $\Hom_\CMs(x,y):=[x,y]$ in $\overline{\CB}$ (or in $\CB$);
\item $\id_x: \one_\CB \to [x,x]$ is the morphism in $\CB$ canonically induced from the unital action $\one_\CB\odot x \simeq x$; 
\item $\circ: [y,z] \otimes [x,y] \to [x,z]$ is the morphism canonically induced from the action $([y,z]\otimes [x,y])\odot x \to [y,z]\odot y \to z$. 
\item $\otimes: [x',y']\otimes [x,y] \to [x'\otimes x, y'\otimes y]$ is the morphism in $\CB$ canonically induced from the action
\begin{align}
([x',y']\otimes [x,y]) \odot x' \otimes x &= \phi_\CM([x',y'] \otimes [x,y]) \otimes x' \otimes x 
\to \phi_\CM([x',y']) \otimes \phi_\CM([x,y]) \otimes x' \otimes x \nn
&\hspace{-1cm} \xrightarrow{\Id \otimes b_{\phi_\CM([x,y]), x'} \otimes \Id_x}  \phi_\CM([x',y']) \otimes x' \otimes \phi_\CM([x,y]) \otimes  x \to y' \otimes y. 
\end{align}
\end{itemize}
\end{exam}


\end{document}